\pdfoutput=1
\documentclass[sigplan,screen,nonacm]{acmart}
\usepackage{asystem-tech-report}

\usepackage{tabularx}
\usepackage{array}
\usepackage{multirow}
\usepackage{pifont}
\usepackage{tikz}
\usepackage{algorithm}
\usepackage{algpseudocode}
\usepackage{enumitem}

\numberwithin{equation}{section}

\DeclareRobustCommand{\cmark}{\textcolor[HTML]{1B7837}{\ding{51}}}
\DeclareRobustCommand{\xmark}{\textcolor[HTML]{B22222}{\ding{55}}}
\DeclareRobustCommand{\pmark}{%
  \tikz[baseline=-0.5ex]{
    \draw[line width=0.55pt, draw=black!75] (0,0) circle (0.62ex);
    \fill[black!75] (0,0) -- (0,0.62ex) arc (90:-90:0.62ex) -- cycle;
  }%
}

\algrenewcommand\algorithmiccomment[1]{\hfill{\footnotesize\textcolor{black!60}{// #1}}}
\algnewcommand{\LineComment}[1]{\Statex {\footnotesize\textbf{\textcolor[HTML]{2B5C8F}{// #1}}}}

\DeclareRobustCommand{\sys}{\textsc{Pacemaker}}

\title[Pacemaker]{Pacemaker: Keeping Training Fed in Long-Horizon Agentic RL via Progress-Driven Training--Rollout Co-Orchestration}

\reportauthor{Songyun Qu}{1}
\thanks{Contact: Songyun Qu (\href{mailto:qusongyun.qsy@antgroup.com}{\texttt{qusongyun.qsy@antgroup.com}}).}
\reportauthor{Yongzhi Xu}{1}
\reportauthor{Tao Wu}{1}
\reportauthor{Yijie Chen}{1}
\reportauthorbreak
\reportauthor{Xiaowei Zhu}{1}
\reportauthor{Zhenxuan Pan}{1}
\reportaffiliation{1}{Ant Group, China}

\begin{document}

\begin{abstract}
Long-horizon agentic reinforcement learning (agentic RL) trains LLMs over multi-turn inference and tool execution. Under group-relative algorithms such as GRPO, wide variance in interaction depth and tool latency creates a \emph{two-level long tail}: a single straggler stalls its entire sampling group, while staggered group completions intermittently starve training. Existing RL systems backfilling cross-step trajectories or shifting GPUs from inference load states alone fail to conserve effective multi-step progress---leaking GPU time into low-batch tail decoding, state recomputation, and barrier stalls because they overlook the \emph{temporal alignment between complete-group supply and training consumption}.

We present \sys{}, a progress-driven training--rollout co-orchestration framework that closes the loop across admission, scheduling, and resource allocation around complete-group delivery (\texttt{Groups/s}). \sys{} integrates: (1)~water-level-gated dynamic admission (L1) to backfill freed slots without KV-cache thrashing and selectively migrate nearly complete groups when rollout GPUs are reclaimed; (2)~group-priority spatio-temporal scheduling (L2) combining group-barrier-aware prioritization, coordinated KV/Mamba state protection, and adaptive prefill batching; and (3)~coverage-driven training--rollout resource allocation (L3) that triggers an initial segmented training phase whenever its execution hides the remaining tail rollout.

Evaluated on a 124B-parameter hybrid linear-attention (KDA/Mamba)--MoE LLM across 64 H200 GPUs, \sys{} delivers 1.30$\times$--1.84$\times$ end-to-end throughput speedups over \texttt{QwenGyre}, \texttt{AReaL}, and \texttt{veRL}, with no convergence degradation.
\end{abstract}

\maketitle

\section{Introduction}
\label{sec:intro}

Long-horizon agentic reinforcement learning (agentic RL) has become a primary driver for advancing large language models (LLMs) on real-world software engineering (e.g., SWE-bench~\cite{jimenez2024swebench}), multi-step reasoning, and complex tool use. Under group-relative policy optimization (e.g., GRPO~\cite{shao2024deepseekmath}), each policy update consumes $B$ complete sampling groups, where each prompt spawns $K$ parallel multi-turn trajectories (all-$K$ completion) that alternate between LLM generation and sandboxed tool execution. Because trajectories diverge across turns in interaction depth, context length, and sandbox latency, variance compounds over time into a severe \emph{two-level long tail} (Figure~\ref{fig:two-level-tail}): \emph{within a group}, one straggler blocks delivery of its $K-1$ finished siblings (e.g., by $3.2\times$ in Figure~\ref{fig:two-level-tail}a); \emph{across groups}, wide arrival dispersion stretches the slowest 25\% of groups across 37.3\% of step rollout time (Figure~\ref{fig:two-level-tail}b). Under a fixed cluster GPU budget, this induces a sharp \emph{supply--demand temporal mismatch}: hundreds of in-flight trajectories occupy GPU compute and HBM capacity without assembling the complete groups needed to unblock the current training step.

To hide these rollout stalls, recent systems rely on two primary mechanisms alongside critical-group scheduling (VenusRL~\cite{zhang2026venusrl}): (i)~\emph{asynchronous training} (e.g., AReaL~\cite{fu2025areal}), which dispatches cross-step requests into inference slots freed by completed trajectories; and (ii)~\emph{dynamic training--rollout resource allocation} (e.g., RollPacker~\cite{gao2026rollpacker}, QwenGyre~\cite{wang2026qwengyre}, DynaRL~\cite{wang2026dynarl}, and Libra~\cite{chen2026libra}), which transitions a subset of rollout GPUs to early training mid-step once partial batches finish (with DynaRL predicting stage reallocations online and Libra co-planning stage ratios with parallelism). However, because group-relative RL requires all-$K$ complete groups under bounded policy staleness, merely backfilling slots or shifting compute from inference load states alone leaves a central systems question unresolved: \emph{how does trajectory-level progress translate into complete-group supply without leaking GPU time across steps?} Inference work not converted into current-step complete groups does not carry over losslessly to accelerate subsequent steps; instead, local metrics diverge from effective \texttt{Groups/s} through three pitfalls:

\begin{figure}[t!]
  \centering
  \includegraphics[width=\columnwidth]{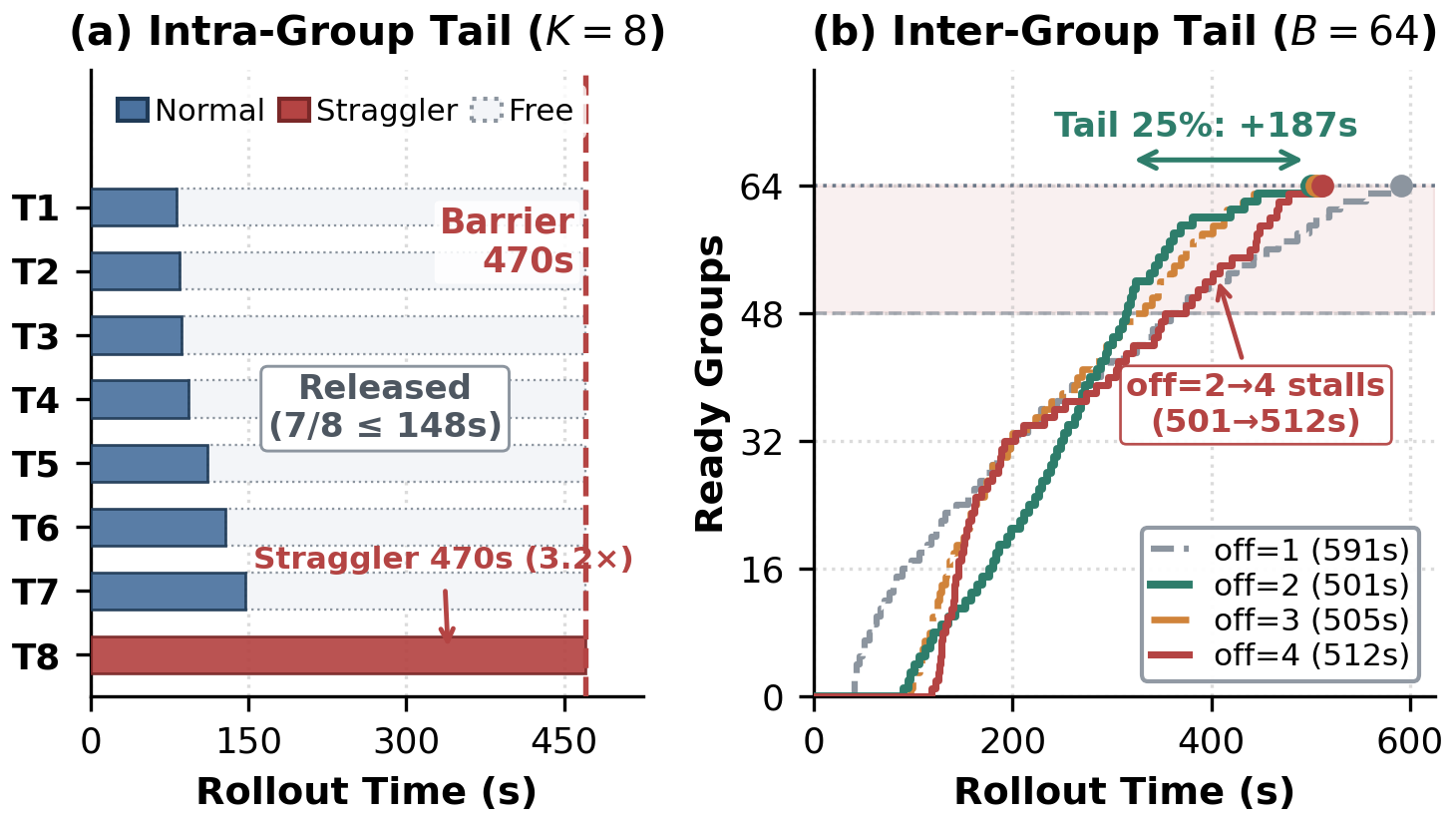}
  \caption{Two-level long-tail phenomenon in agentic RL: (a)~intra-group straggler barrier ($K=8$); (b)~inter-group tail across in-flight batch windows (\texttt{offpolicy} $\in \{1..4\}$, $B=64$).}
  \label{fig:two-level-tail}
\end{figure}

\begin{itemize}[leftmargin=*, topsep=2pt, itemsep=1.5pt]
  \item \textbf{Cross-step backfilling cannot boundlessly fill tails or avoid recomputation.} Bounded staleness caps lookahead once stragglers stall weight updates; meanwhile, exceeding the physical KV water-level triggers KV/Mamba eviction recomputation that slows active stragglers, whereas co-located full-batch switching purges in-flight states at step boundaries and pays heavy re-prefills next step.

  \item \textbf{Per-trajectory or KV-only group scheduling fails to prioritize and shield nearly complete groups.} Under an intra-group barrier, scheduling requests by arrival order or single-trajectory turn counts (e.g., DynaRL~\cite{wang2026dynarl}) spreads decode slots across partial groups far from completion. Meanwhile, existing critical-group schedulers (e.g., VenusRL~\cite{zhang2026venusrl}) only pin standard KV caches, leaving hybrid KV+Mamba states vulnerable to eviction and allowing incoming prefills to stall tail decoding.

  \item \textbf{Mid-step reallocation is essential under bounded staleness, yet inference-only triggers cause imbalance.} Static disaggregation idles trainer GPUs at slow-step barriers, whereas keeping all GPUs in rollout once staleness caps out and stragglers wait on CPU sandboxes strands MoE engines in bandwidth-bound low-batch decoding. Shifting excess rollout GPUs to early training converts under-utilized tail GPU-seconds into updates and unlocks the next staleness window sooner, yet existing designs observe only inference states: utilization-decay triggers (RollPacker~\cite{gao2026rollpacker}, DynaRL~\cite{wang2026dynarl}) vanish under cross-step traffic, while slot and preset-ratio triggers (QwenGyre~\cite{wang2026qwengyre}) ignore whether ready-group training matches remaining tail rollout---switching too early (starving training) or too late (wasting overlap).
\end{itemize}

These pitfalls reveal a core systems principle: \emph{dynamic request admission, in-flight priority scheduling, and training--rollout resource allocation are not independent local knobs, but a tightly coupled control loop centered on the complete-group supply process}. Admission sets the physical KV water-level and contention boundary; scheduling governs whether bounded compute converts cleanly into complete groups without state eviction or prefill stalls; and the resulting group-arrival rate dictates when training--rollout resource allocation should launch early segmented training to mask post-switch tail rollout.

Guided by this principle, we design and build \sys{}, a \emph{progress-driven training--rollout co-orchestration framework} for long-horizon agentic RL that paces complete-group supply to match training consumption across three coupled levels. \textbf{(1)~Water-level-gated dynamic request admission (L1, \S\ref{sec:method-l1}).} As fast trajectories finish and release slots (e.g., 7 of 8 done), L1 prioritizes current-step and warm sticky continuations ahead of cross-step prefetches, enforcing a physical KV water-level gate with KV-balanced replica placement in SGLang~\cite{zheng2024sglang} and selectively migrating high-completeness groups upon elastic role switching to prevent post-switch cache thrashing. \textbf{(2)~Group-priority spatio-temporal adaptive scheduling (L2, \S\ref{sec:method-l2}).} \sys{} derives dynamic priorities from group-completeness barrier sensitivity, jointly governing request queue order and coordinated co-eviction across KV-cache pages and hybrid linear-attention (KDA/Mamba) recurrent state slots to shield nearly complete groups while adapting cross-instance prefill batch sizes and deferral windows to high-priority decode pressure. \textbf{(3)~Coverage-driven training--rollout resource allocation (L3, \S\ref{sec:method-l3}).} Leveraging segmented gradient accumulation to launch an initial training phase (Stage-1) before full-batch rollout completes, \sys{} feeds the L1/L2-shaped group arrivals into a dual-side online predictor that compares ready-group Stage-1 training time against post-switch tail rollout time, switching GPUs at the exact coverage crossing point.

\begin{figure*}[t!]
  \centering
  \includegraphics[width=\textwidth]{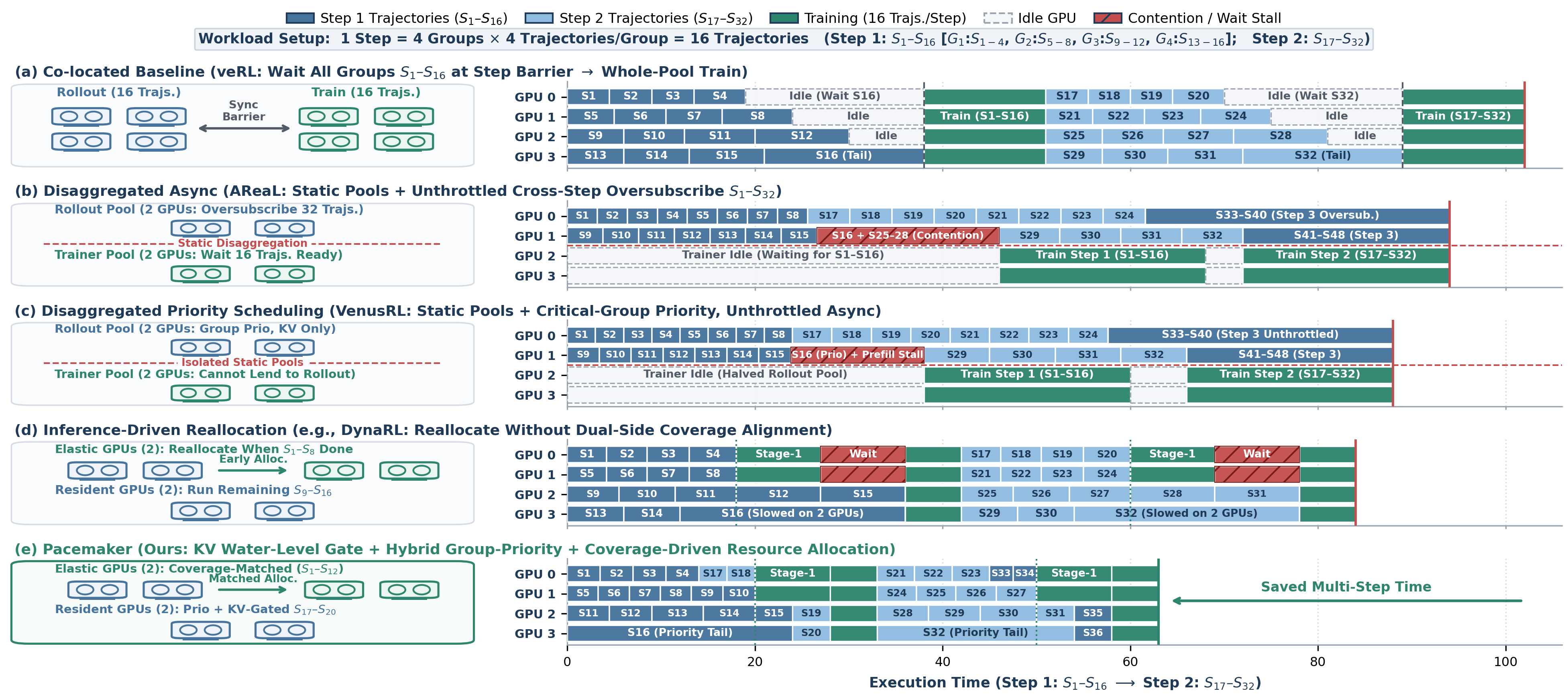}
  \vspace{-5pt}
  \caption{Execution timelines of training--rollout orchestration paradigms across two steps on 4 GPUs ($B=4$ groups/step, $K=4$ trajectories/group, $S_1\text{--}S_{32}$).}
  \vspace{-3pt}
  \label{fig:paradigm-comparison}
\end{figure*}

This paper makes three primary contributions:
\begin{enumerate}[leftmargin=*, topsep=2pt, itemsep=1.5pt]
  \item \textbf{Problem characterization and closed-loop insight.} We show why concurrency expansion, local token throughput, and inference-only load triggers diverge from effective \texttt{Groups/s} in long-horizon agentic RL, motivating closed-loop admission, scheduling, and training--rollout resource allocation.
  \item \textbf{Progress-driven three-level framework.} We present \sys{}, uniting physical KV water-level admission with selective migration (L1), barrier-sensitivity scheduling with hybrid KV/Mamba state protection and adaptive prefill batching (L2), and coverage-driven training--rollout resource allocation (L3).
  \item \textbf{End-to-end evaluation and convergence fidelity.} On 64 NVIDIA H200 GPUs running the open-source 124B hybrid KDA/Mamba--MoE LLM (\texttt{Ling-3.0-Flash}~\cite{inclusionai2026ling3flash}), \sys{} achieves 1.30$\times$--1.84$\times$ end-to-end throughput speedups over \texttt{QwenGyre}, \texttt{AReaL}, and \texttt{veRL}, with no convergence degradation on SWE-Pro~\cite{deng2026swebenchpro} and SWE-ML~\cite{yang2025swesmith}.
\end{enumerate}

\section{Background and Motivation}
\label{sec:background}

\begin{table*}[t!]
  \centering
  \footnotesize
  \caption{Comparison of representative RL post-training systems and \sys{} across four core dimensions (\cmark: full; \pmark: partial; \xmark: none).}
  \vspace{-5pt}
  \label{tab:system-comparison}
  \renewcommand{\arraystretch}{1.12}
  \setlength{\tabcolsep}{1.5pt}
  \begin{tabular*}{\textwidth}{@{\extracolsep{\fill}}l l l l l@{}}
    \toprule
    \textbf{System} &
    \textbf{Async \& Admission} &
    \textbf{Group \& State Sched.} &
    \textbf{Train--Rollout Allocation} &
    \textbf{Static / Dynamic Alloc.} \\
    \midrule
    \textbf{AReaL}~\cite{fu2025areal} &
    \pmark\ Unthrottled cross-step &
    \xmark\ Group-agnostic FCFS &
    \xmark\ Isolated pools &
    \xmark\ Static fixed split \\
    \textbf{RollPacker}~\cite{gao2026rollpacker} &
    \xmark\ Sync oversubscription &
    \xmark\ Aborts slow trajectories &
    \cmark\ Streaming 50\% GPU shift &
    \pmark\ Static ratio + KV cap \\
    \textbf{VenusRL}~\cite{zhang2026venusrl} &
    \pmark\ Unthrottled cross-step &
    \pmark\ Group prio, KV pin only &
    \xmark\ Isolated pools &
    \xmark\ Static fixed split \\
    \textbf{QwenGyre}~\cite{wang2026qwengyre} &
    \pmark\ Slot-capped async &
    \xmark\ Per-request FCFS &
    \cmark\ Cell-wise streaming &
    \pmark\ Dynamic (slot + $\rho_{\min}$) \\
    \textbf{DynaRL}~\cite{wang2026dynarl} &
    \xmark\ Single-step sync &
    \pmark\ Per-traj.\ turn/len prio &
    \cmark\ Stage GPU reallocation &
    \pmark\ Dynamic (inference KV) \\
    \midrule
    \textbf{\sys{}} &
    \textbf{\cmark\ KV water-level gated} &
    \textbf{\cmark\ Group prio, state \& prefill} &
    \textbf{\cmark\ Realloc.\ + selective mig.} &
    \textbf{\cmark\ Dynamic (dual-side align)} \\
    \bottomrule
  \end{tabular*}
  \vspace{-3pt}
\end{table*}

\begin{figure*}[t!]
  \centering
  \includegraphics[width=\textwidth]{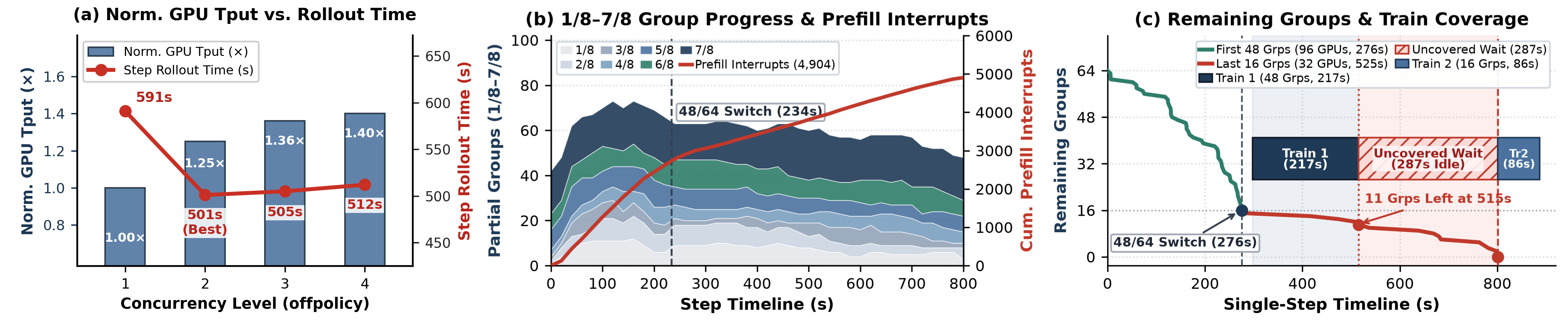}
  \vspace{-5pt}
  \caption{Empirical profiling of 124B Ling-3.0-Flash on 96 H20 GPUs ($B=64\times 8$, representative profiled steps across panels): (a)~rollout time vs.\ in-flight batch window (\texttt{offpolicy}); (b)~partial groups and prefill-induced decode stalls; (c)~inference-only allocation tail wait.}
  \vspace{-3pt}
  \label{fig:motivation-profiling}
\end{figure*}

\subsection{Group Synchronization Barriers and Two-Level Tail Amplification}
\label{sec:bg-tail}

In long-horizon agentic RL, trajectories alternate between LLM generation turns and sandboxed tool execution (shell commands, code edits, and unit tests) with growing context. Under GRPO~\cite{shao2024deepseekmath}, each update (Step~$j$) consumes $B$ complete groups of $K$ parallel trajectories $(t_{g,1}, \dots, t_{g,K})$, inducing a strict \emph{two-level long tail} (Figure~\ref{fig:two-level-tail}):

\begin{itemize}[leftmargin=*, topsep=1.5pt, itemsep=1pt]
  \item \textbf{Level 1: Intra-group barrier and released slots (Figure~\ref{fig:two-level-tail}a).} Group advantage normalization requires terminal rewards from all $K$ members ($p_g(T) = [F(T)]^K$ under i.i.d.\ completion CDF $F(T)$). In Figure~\ref{fig:two-level-tail}a ($K=8$), \texttt{T1--T7} finish within 148\,s and free slots (\texttt{Released Slots}), while one 470\,s ($3.2\times$) straggler (\texttt{T8}) stalls group delivery to $T_g = 470\text{\,s}$.

  \item \textbf{Level 2: Inter-group arrival dispersion (Figure~\ref{fig:two-level-tail}b).} Across $B=64$ groups, the first 48 complete groups (75\%) arrive within 314\,s, whereas the slowest 25\% ($48\to 64$) add 187\,s ($314\text{\,s}\to 501\text{\,s}$, or 37.3\% of rollout time), leaving completed groups idling behind the batch barrier before policy updates can begin.
\end{itemize}

\subsection{Existing Approaches and Their Limitations}
\label{sec:bg-related}

Prior systems target three directions (Figure~\ref{fig:paradigm-comparison}, Table~\ref{tab:system-comparison}):

\begin{itemize}[leftmargin=*, topsep=1.5pt, itemsep=1pt]
  \item \textbf{Async cross-step sampling and sync tail batching.} Co-located cross-step carry-over (e.g., async \texttt{veRL}~\cite{sheng2025hybridflow}) purges GPU KV/Mamba caches whenever all GPUs switch to training, forcing carried-over long-context trajectories to pay expensive prefix re-prefills on the next step. AReaL~\cite{fu2025areal} avoids state purges by statically partitioning trainer and rollout pools with staleness-bounded admission, yet without group-barrier awareness, contention gating, or cross-pool GPU sharing, group barriers leave trainer GPUs idling on heavy-tail steps and rollout GPUs idling on light-tail steps. RollPacker~\cite{gao2026rollpacker} oversubscribes synchronous ``short rounds'' ($\eta=1.25$) and aborts slow requests to replay in ``long rounds'' while streaming half of rollout GPUs to training at 20\%--50\% completion; yet aborting stateful multi-turn sandbox trajectories (e.g., containerized edits and shell sessions) is prohibitively costly, and request-fraction triggers ignore group completion.

  \item \textbf{Critical-group scheduling under static disaggregation.} Under unthrottled cross-step dispatch, VenusRL~\cite{zhang2026venusrl} uses a turn-boundary length predictor to assign binary critical-group priorities for slot preemption, Radix KV retention, and worker migration. Yet static disaggregation cannot borrow training GPUs mid-step, lacks admission gating, and neither coordinates non-sliceable recurrent KDA/Mamba states in hybrid models~\cite{deepseek2024v2,gu2024mamba} (whose fixed-size state slots must be co-evicted alongside paged KV blocks) nor mitigates scattered-prefill stalls under MoE expert parallelism (EP).

  \item \textbf{Inference-driven training--rollout resource management.} QwenGyre~\cite{wang2026qwengyre} transitions ordered elastic cells serially into streaming 1F1B training once remaining rollout slots cover unfinished executions ($w(t) \le C_R(t) - C_e$) and ready groups reach a preset ratio ($\rho_{\min} B$). DynaRL~\cite{wang2026dynarl} prioritizes per-trajectory turns and length $(k_r, m_r)$ and triggers hypergraph reallocation on single-step KV under-utilization ($U_D(t) < U_{\mathrm{low}}$); Libra~\cite{chen2026libra} co-plans stage ratios with parallelism, and ReFlow~\cite{tian2026reflow} reallocates GPUs inter-step. Crucially, \emph{these designs allocate training--rollout compute solely from inference-side load states}: utilization-decay triggers (RollPacker, DynaRL) vanish when asynchronous cross-step traffic keeps slots saturated, while execution-count waterlevels and preset group ratios ($\rho_{\min}$ in QwenGyre) ignore whether ready-group training duration matches remaining tail rollout. \sys{} instead bounds async admission concurrency, prioritizes and shields nearly complete groups across hybrid states, and aligns training--rollout allocation via dual-side supply--demand prediction.
\end{itemize}

\subsection{Empirical Motivation and Design Insights}
\label{sec:bg-motivation}

Diagnostic profiling of the open-source 124B Ling-3.0-Flash~\cite{inclusionai2026ling3flash} hybrid model on SWE agent tasks ($B=64$, $K=8$) on a 96-GPU H20 cluster (64 elastic + 32 resident GPUs) in Figure~\ref{fig:motivation-profiling} reveals three insights:

\paragraph{1) Admission water-level boundary (Figure~\ref{fig:motivation-profiling}a).}
Raising the active in-flight batch window (\texttt{offpolicy}, where \texttt{1} denotes synchronous single-step batching and $W>1$ allows up to $W-1$ cross-step lookahead steps) from \texttt{1} to \texttt{2} reuses freed slots, lifting normalized GPU throughput by 1.25$\times$ and cutting step rollout time from 591\,s to 501\,s. Pushing \texttt{offpolicy} to \texttt{3} and \texttt{4} raises raw normalized throughput to 1.36$\times$--1.40$\times$ on next-step prefetch work while intensifying KV and KDA/Mamba contention, causing current-step completion to stall (505\,s $\to$ 512\,s). Cross-step prefetch admission must thus be gated at a physical KV water-level.

\paragraph{2) Group fragmentation and prefill interference (Figure~\ref{fig:motivation-profiling}b).}
Under an $8$-member barrier, advancing eight groups to $7/8$ (56 trajectories) yields 0 trainable groups whereas finishing seven (56 trajectories) yields 7; yet group-agnostic scheduling lets backfilled requests split decode slots and trigger LRU eviction of KV and KDA/Mamba state for nearly complete stragglers. Meanwhile, uncoordinated turn restarts launch heavy prefills during tail decode, accumulating partial groups (\texttt{1/8--7/8}) and inflicting 4,904 prefill-induced decode interruptions per step. Even with chunked prefill~\cite{agrawal2024sarathi} enabled, each prefill chunk on any replica still forces peer decode replicas to wait at synchronous MoE All-to-All barriers and recurrent KDA/Mamba state-scan kernels, while P--D disaggregation~\cite{zhong2024distserve} fragments HBM/EP capacity and incurs cross-pool state transfers on every tool return. Sustaining high \texttt{Groups/s} thus requires barrier-aware hybrid state protection and cross-replica prefill batching.

\paragraph{3) Inference-only allocation imbalance (Figure~\ref{fig:motivation-profiling}c).}
Reallocating 64 elastic GPUs to an initial segmented training phase (\texttt{Train~1}) at a fixed inference-side threshold of $q=48$ ready groups ($t=276\text{\,s}$) leaves 32 resident GPUs to finish the remaining 16 groups, stretching post-switch tail rollout duration to 525\,s (completing at $t \approx 801\text{\,s}$). After a 22\,s switching overhead, \texttt{Train~1} finishes in 217\,s ($t=515\text{\,s}$) with 11 groups still incomplete, forcing 287\,s of uncovered trainer stall before final-segment training (\texttt{Tr2}, 86\,s).

\paragraph{Takeaway.}
Cross-step backfilling and inference-only triggers do not conserve GPU time across steps when work leaks into low-batch tail decoding, prefix/eviction recomputation, and barrier stalls; keeping training fed in long-horizon agentic RL instead requires \emph{closed-loop supply--demand pacing} across admission, group scheduling, and training--rollout resource allocation.

\section{Pacemaker Design}
\label{sec:method}

\begin{figure*}[t!]
  \centering
  \includegraphics[width=\textwidth]{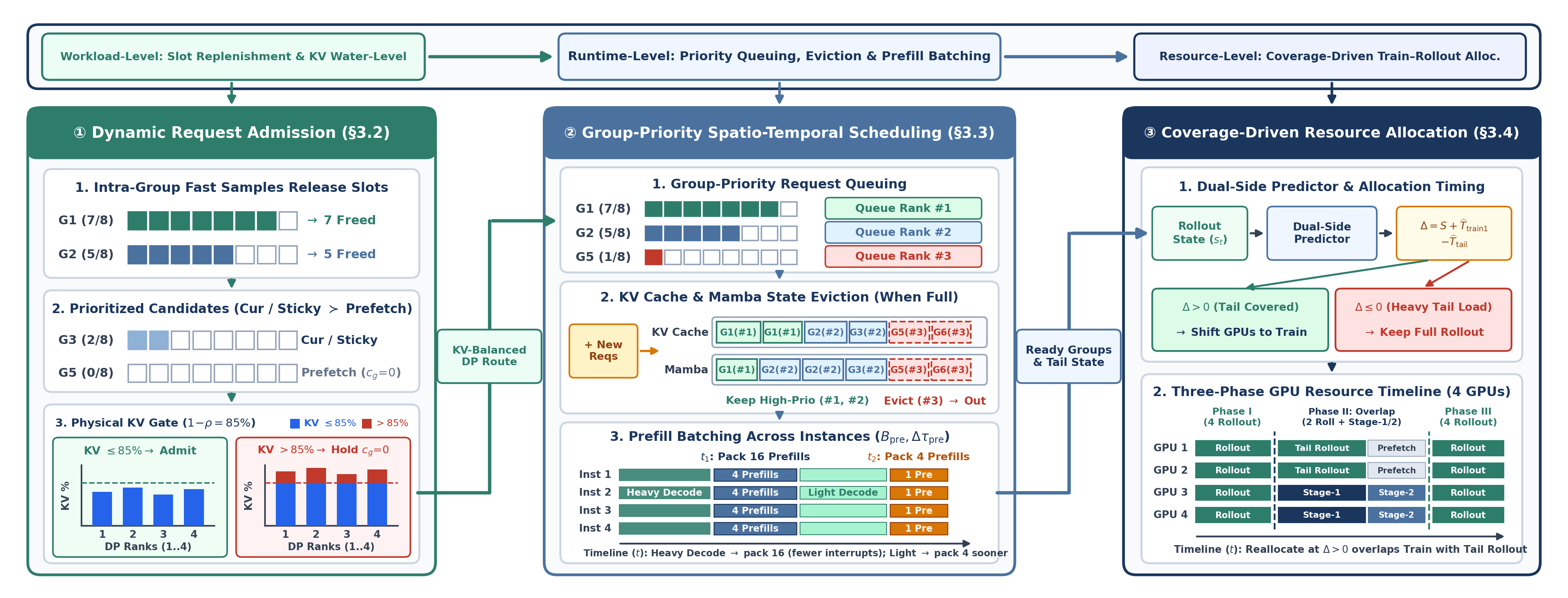}
  \caption{Overview of \sys{}'s three-level co-orchestration architecture: (1)~L1 dynamic request admission (\S\ref{sec:method-l1}), (2)~L2 group-priority scheduling (\S\ref{sec:method-l2}), and (3)~L3 coverage-driven training--rollout resource allocation (\S\ref{sec:method-l3}).}
  \label{fig:overview}
\end{figure*}

Guided by the unified objective of pacing complete-group supply to match training consumption and minimizing update cycle time, \sys{} organizes trajectory generation and training into a closed-loop \emph{Three-Level Orchestration Hierarchy} (Figure~\ref{fig:overview}):
(1)~\emph{Workload-Level (L1, \S\ref{sec:method-l1})} dynamically backfills freed inference slots under a physical KV water-level gate (prioritizing current-step and warm sticky continuations over cross-step prefetch) with KV-balanced replica placement and selective high-completeness migration upon elastic role switching;
(2)~\emph{Runtime-Level (L2, \S\ref{sec:method-l2})} schedules in-flight trajectories by group-completeness barrier sensitivity, jointly governing queue order, hybrid KV+Mamba co-eviction, and load-aware prefill batching; and
(3)~\emph{Resource-Level (L3, \S\ref{sec:method-l3})} predicts Stage-1 training versus post-switch tail rollout durations from live group arrivals, reallocating GPUs from rollout to training at the exact coverage crossing point.

\subsection{Execution Model and Global State Abstraction}
\label{sec:method-model}

Consider a cluster of $N$ GPUs executing group-relative RL, where each parameter update (Step~$j$) consumes $B$ complete sampling groups of $K$ multi-turn trajectories. Within each step, the cluster cycles through three execution phases (Figure~\ref{fig:overview}, bottom right):
(1)~\emph{Phase~I (Full Rollout)}, where all $N$ GPUs run rollout generation to rapidly accumulate complete groups;
(2)~\emph{Phase~II (Partial Rollout + Train)}, where once $q$ complete groups arrive at the switch point, $N-n$ elastic GPUs transition to Stage-1 gradient computation while $n$ resident GPUs finish the remaining $B-q$ current-step tail groups (plus cross-step prefetches), followed on those $N-n$ trainer GPUs by Stage-2 gradient computation and the optimizer step once all $B$ groups finish; and
(3)~\emph{Phase~III (Back to Full Rollout)}, where updated weights broadcast to all replicas and all $N$ GPUs resume full rollout for Step~$j+1$.

To guarantee exact mathematical equivalence between early-started Phase~II training and synchronous full-batch updates, \sys{} exploits fixed within-step policy parameters $\theta$ via \emph{segmented gradient accumulation}. Partitioning $\mathcal{B}_j = \mathcal{B}_{j,1} \sqcup \mathcal{B}_{j,2}$ into the first $q$ ready groups and trailing $B-q$ groups, each segment runs forward--backward passes under $\theta$ while accumulating unnormalized group loss numerators $L_g^{\mathrm{sum}}(\theta)$ and segment valid-token counts $N_{\mathrm{tok},s}$ ($N_{\mathrm{tok}} = \sum_{s=1}^{2} N_{\mathrm{tok},s}$):
\begin{equation}
\begin{aligned}
  L_j(\theta) &= \frac{1}{N_{\mathrm{tok}}} \sum_{g \in \mathcal{B}_j} L_g^{\mathrm{sum}}(\theta), \\
  \nabla L_j(\theta) &= \frac{1}{\sum_{s=1}^{2} N_{\mathrm{tok},s}} \sum_{s=1}^{2} \sum_{g \in \mathcal{B}_{j,s}} \nabla L_g^{\mathrm{sum}}(\theta),
\end{aligned}
\label{eq:seg-grad}
\end{equation}
followed by a single global normalization, gradient clipping, and optimizer step. Because the $n$ resident GPUs continue tail rollout while the $N-n$ elastic GPUs undergo role-switching handoff $S(q)$, the per-step update cycle time $T_j(q)$ decomposes as:
\begin{equation}
\begin{aligned}
  T_j(q) &= T_{\mathrm{roll}}(q) + \max\!\big\{S(q) + T_{\mathrm{train1}}(q),\, T_{\mathrm{tail}}(q)\big\} \\
         &\quad + T_{\mathrm{train2}}(q) + H \\
         &= T_{\mathrm{roll}}(q) + S(q) + T_{\mathrm{train}} \\
         &\quad + \big[T_{\mathrm{tail}}(q) - \big(S(q) + T_{\mathrm{train1}}(q)\big)\big]_+ + H,
\end{aligned}
\label{eq:step-decomp}
\end{equation}
where $T_{\mathrm{roll}}(q)$ is Phase~I time for the first $q$ groups on $N$ GPUs, $T_{\mathrm{train1}}(q) + T_{\mathrm{train2}}(q) = T_{\mathrm{train}}$ is total training time on $N-n$ GPUs, $T_{\mathrm{tail}}(q)$ is Phase~II time for $n$ resident GPUs to finish the remaining $B-q$ groups, and $H$ is optimizer and weight-broadcast overhead. Total \emph{exposed rollout time} (\S\ref{sec:eval}) is $T_{\mathrm{roll}}(q) + \big[T_{\mathrm{tail}}(q) - (S(q) + T_{\mathrm{train1}}(q))\big]_+$ (Phase~I rollout plus any uncovered post-switch tail wait). Equation~\eqref{eq:step-decomp} captures our cross-level synergy: L1 and L2 shrink $T_{\mathrm{roll}}(q)$ and $T_{\mathrm{tail}}(q)$, while L3 switches once $\Delta(q) = S(q) + \widehat{T}_{\mathrm{train1}}(q) - \widehat{T}_{\mathrm{tail}}(q) > 0$, eliminating uncovered post-switch tail wait.

\paragraph{Global runtime state abstraction $s_t$.}
To drive closed-loop control across all three levels, \sys{} continuously tracks three lightweight runtime states in $s_t$:
(1)~\emph{group progress $(c_g, a_g, u_g)$}, recording the completed ($c_g$), active ($a_g$), and pending ($u_g$) trajectory counts of each group~$g$ ($c_g + a_g + u_g = K$), where group~$g$ enters the ready queue $\mathcal{Q}_j^{\mathrm{ready}}$ once $c_g = K$;
(2)~\emph{replica load $(\mathrm{Run}_r, M_r^{\mathrm{used}}, M_r^{\mathrm{transit}}, C_r^{\mathrm{KV}})$}, capturing active running requests, used and in-transit KV token footprints, and KV capacity (alongside Mamba slot occupancy) on each replica $r \in \{1,\dots,R\}$; and
(3)~\emph{ready-group count $q = |\mathcal{Q}_j^{\mathrm{ready}}|$} along with its cumulative valid token count $Q_j(q)$.

\subsection{Workload-Level (L1): Water-Level-Gated Dynamic Request Admission}
\label{sec:method-l1}

\begin{figure}[t!]
  \centering
  \includegraphics[width=\columnwidth]{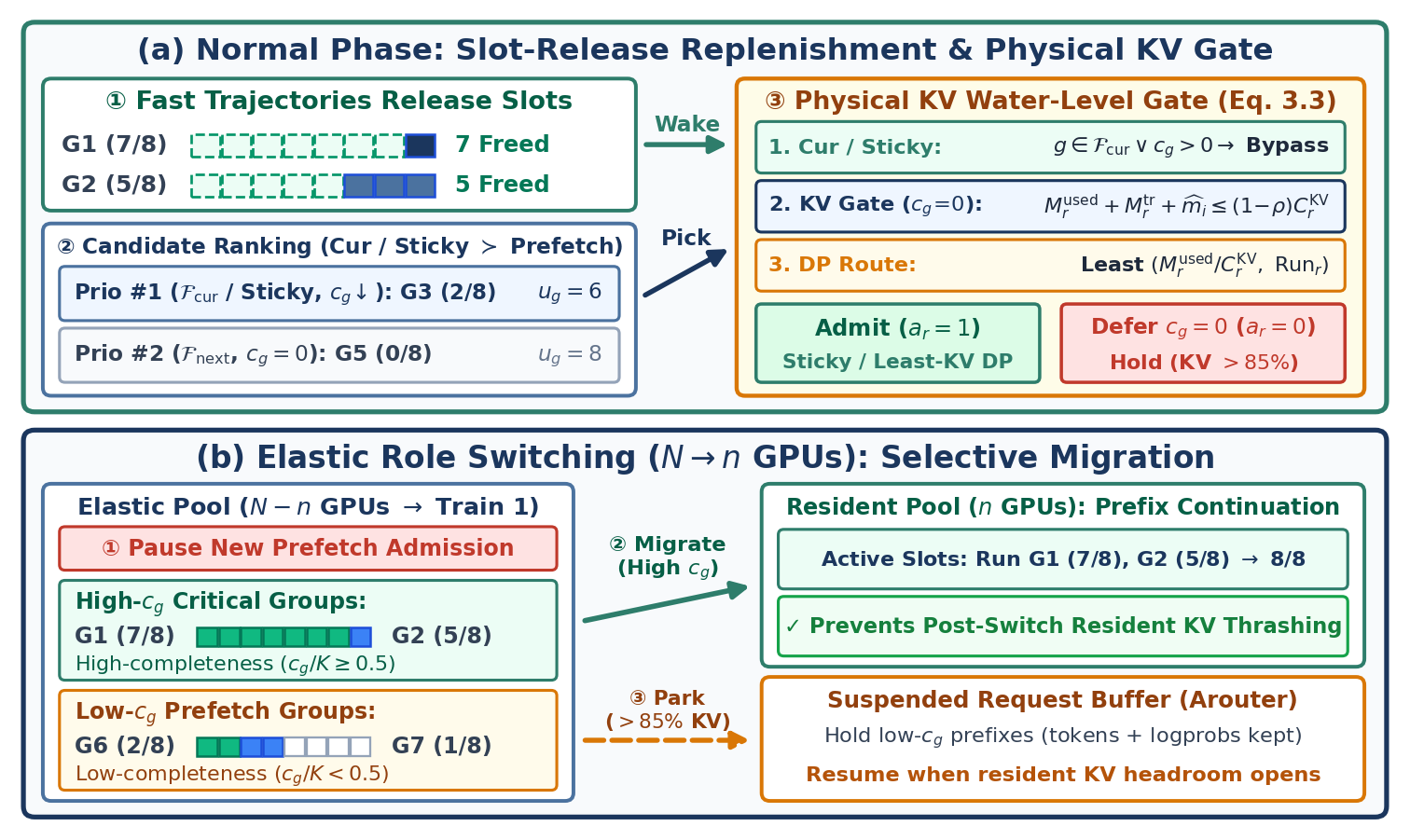}
  \caption{Workload-level (L1) dynamic admission: (a)~slot-release replenishment and physical KV water-level gating; (b)~selective migration upon elastic role switching ($N \to n$ GPUs).}
  \label{fig:l1-admission}
\end{figure}

In fixed-batch rollout, slots freed by fast siblings sit idle waiting on intra-group stragglers, whereas un-gated off-policy oversubscription floods replica KV pools, triggering heavy cache thrashing and decode slowdown (Figure~\ref{fig:motivation-profiling}a). \sys{} resolves this tension via \emph{slot-release wakeup $\to$ prioritized candidate ranking $\to$ physical KV water-level gating, balanced DP placement, and selective migration} (Figure~\ref{fig:l1-admission}).

\paragraph{1. Slot-release triggering and lexicographic candidate prioritization.}
Whenever a trajectory finishes a turn or terminates, L1 wakes immediately to dispatch waiting requests. L1 orders candidates lexicographically:
(1)~\emph{Priority~1 (Current-step groups $\mathcal{F}_{\mathrm{cur}}$ and warm sticky continuations)}, ranked by warm KV affinity and descending $c_g$ to complete nearly finished groups first (e.g., \texttt{G3 (2/8)} in Figure~\ref{fig:overview}); and
(2)~\emph{Priority~2 (Brand-new cross-step prefetch trajectories $g \in \mathcal{F}_{\mathrm{next}},\, c_g = 0$)} (e.g., \texttt{G5 (0/8)}), admitted within the staleness budget only when physical KV headroom permits.

\paragraph{2. Physical KV water-level gating, balanced placement, and selective migration.}
Because cross-node scheduler telemetry is sampled asynchronously while SGLang's in-process L2 queue manages compute batching at sub-millisecond granularity, locking replicas on coarse queue-depth snapshots creates artificial dispatch bubbles. Instead, L1 admits all current-step requests and warm multi-turn continuations without artificial queueing locks, and enforces a \emph{physical KV water-level gate} (headroom margin $\rho$, default $\rho=0.15$, i.e., 85\% KV water-level) exclusively on brand-new cross-step prefetches ($g \in \mathcal{F}_{\mathrm{next}},\, c_g = 0$; Figure~\ref{fig:l1-admission}a). Let $M_r^{\mathrm{used}}$, $M_r^{\mathrm{transit}}$, and $C_r^{\mathrm{KV}}$ denote replica~$r$'s used KV tokens, in-transit reserved tokens, and KV token capacity, and $\widehat{m}_i$ denote candidate~$i$'s initial KV footprint. Replica~$r$ is eligible ($a_r=1$) iff Mamba state capacity holds and:
\begin{equation}
\begin{aligned}
  &a_r = 1 \iff \big(g \in \mathcal{F}_{\mathrm{cur}} \lor \mathrm{sticky}(i) \lor c_g > 0\big) \\
  &\quad\lor \big(M_r^{\mathrm{used}} + M_r^{\mathrm{transit}} + \widehat{m}_i \le (1 - \rho)C_r^{\mathrm{KV}}\big).
\end{aligned}
\label{eq:admission-gate}
\end{equation}
Multi-turn continuations preserve warm sticky affinity to their prefix replica; initial first-turn candidates without replica affinity are placed on the least-loaded eligible DP rank ordered by $(M_r^{\mathrm{used}}/C_r^{\mathrm{KV}},\allowbreak\, \mathrm{Run}_r,\allowbreak\, M_r^{\mathrm{bound}})$ (with $M_r^{\mathrm{bound}}$ tracking sticky-bound tokens) to prevent KV hotspots. When Phase~I transitions to Phase~II ($N \to n$ rollout GPUs), L1 migrates all unfinished current-step trajectories ($g \in \mathcal{F}_{\mathrm{cur}}$, ordered by descending $c_g$) along with high-completeness prefetches ($c_g / K \ge 0.5$) within resident headroom, while parking low-completeness cross-step prefetches ($g \in \mathcal{F}_{\mathrm{next}},\, c_g / K < 0.5$) whenever resident KV utilization reaches $(1-\rho)$ (Figure~\ref{fig:l1-admission}b) to prevent post-switch cache thrashing without stalling Step~$j$ while keeping resident prefixes warm across step boundaries. Still, under bounded staleness and sandbox waits, admission gating alone prevents neither prefetch dilution of nearly complete groups (L2) nor low-batch tail under-utilization once lookahead saturates (L3).

\subsection{Runtime-Level (L2): Group-Priority Spatio-Temporal Adaptive Scheduling}
\label{sec:method-l2}

\begin{figure}[t!]
  \centering
  \includegraphics[width=\columnwidth]{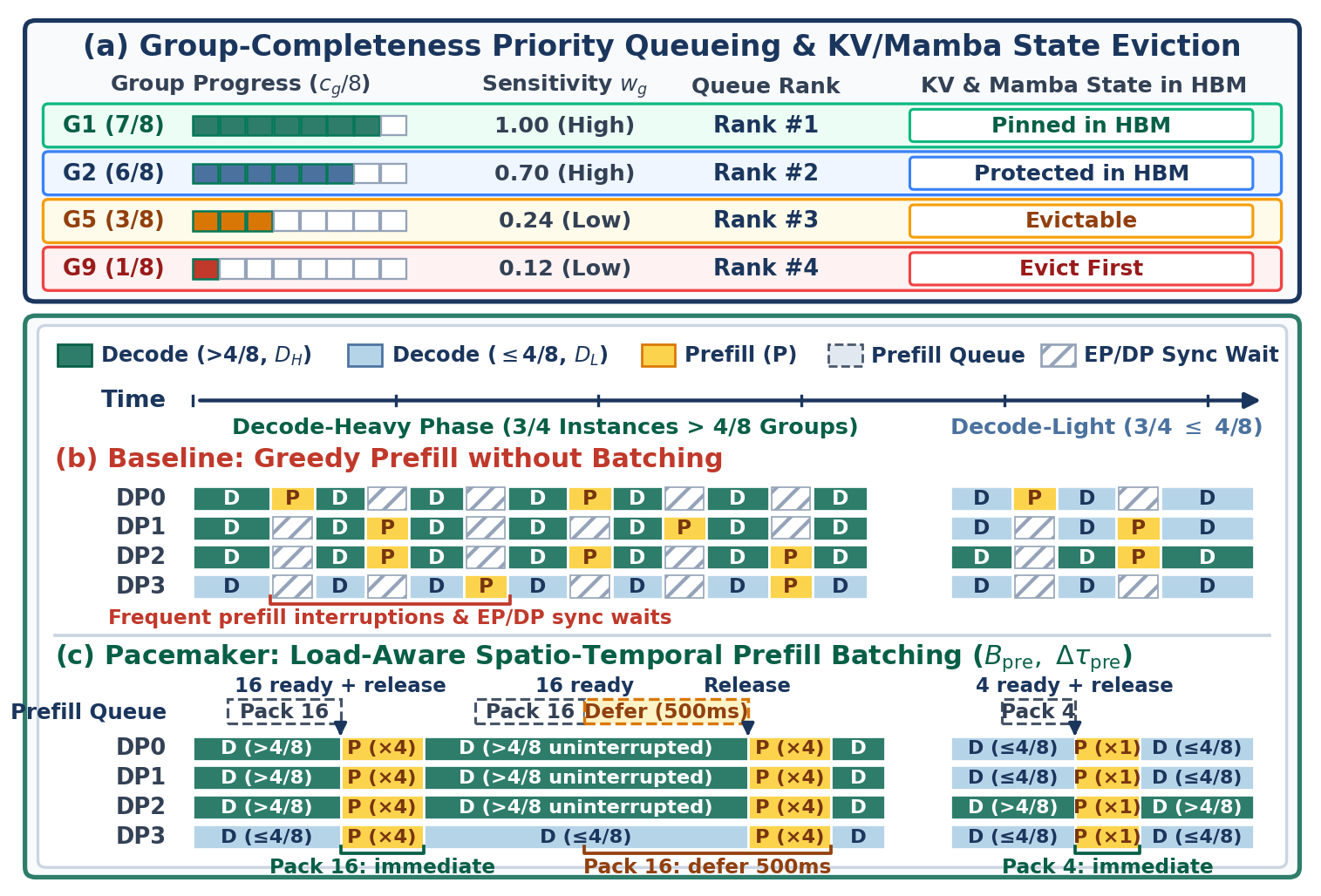}
  \caption{Runtime-level (L2) scheduling: (a)~group-completeness priority ranking and KV/Mamba state co-eviction; (b)~unbatched greedy prefill stalls; (c)~load-aware spatio-temporal prefill batching.}
  \label{fig:l2-scheduling}
\end{figure}

Even under a physical KV water-level, group-agnostic FCFS queueing, LRU eviction, and greedy immediate prefills allow low-progress backfilled requests (\texttt{1/8}) to contend with nearly complete stragglers (\texttt{7/8}) and interrupt decode batches (Figure~\ref{fig:motivation-profiling}b). L2 co-designs \emph{request queueing}, \emph{hybrid KV/Mamba eviction}, and \emph{cross-instance prefill batching} around group barrier sensitivity (Figure~\ref{fig:l2-scheduling}).

\paragraph{1. Barrier sensitivity and group-priority request queueing.}
For an incomplete group $g$ with unfinished set $\mathcal{U}_g$ ($|\mathcal{U}_g| = K - c_g$) and per-trajectory completion CDF $F_\ell(T) \in (0,1)$ within window~$T$, group readiness probability is $p_g(T) = \prod_{\ell \in \mathcal{U}_g} F_\ell(T)$. Differentiating with respect to $F_i(T)$ yields trajectory~$i$'s marginal \emph{barrier sensitivity} $w_g$:
\begin{equation}
  w_g = \frac{\partial p_g(T)}{\partial F_i(T)}
  = \prod_{\ell \in \mathcal{U}_g \setminus \{i\}} F_\ell(T)
  \approx \bar{F}(T)^{\,K - c_g - 1}.
  \label{eq:barrier-sens}
\end{equation}
Because $0 < F_\ell(T) < 1$, fewer remaining members $K - c_g$ exponentially amplify marginal group-delivery gain: at $\bar{F}(T)=0.7$ (Figure~\ref{fig:l2-scheduling}a), \texttt{G1 (7/8)} and \texttt{G2 (6/8)} reach sensitivities $0.7^0 = 1.00$ and $0.7^1 = 0.70$, versus $0.7^4 \approx 0.24$ and $0.7^6 \approx 0.12$ for \texttt{G5 (3/8)} and \texttt{G9 (1/8)}. \sys{} defines runtime priority score $\Pi(i)$ for trajectory $i \in g$ as:
\begin{equation}
  \Pi(i) = \lambda_{\mathrm{step}}\,\mathbb{I}\big(g \in \mathcal{F}_{\mathrm{cur}}\big)
  + \frac{w_g}{\widehat{d}_i},
  \label{eq:priority-score}
\end{equation}
where $\lambda_{\mathrm{step}} \gg 1$ prioritizes current-step groups over cross-step prefetches and $\widehat{d}_i$ estimates remaining turn decode time from EMA-smoothed turn token lengths and replica decode throughput (analogous to shortest-effective-bottleneck coflow prioritization~\cite{chowdhury2014varys,chowdhury2015aalo}), ordering replica queues by descending $\Pi(i)$ (\texttt{Rank \#1 $\to$ \#4}).

\paragraph{2. Priority-aware KV Cache and KDA/Mamba state co-eviction.}
Hybrid architectures combining MLA~\cite{deepseek2024v2} with linear attention (KDA~\cite{kimi2025linear}/Mamba~\cite{gu2024mamba}) hold two heterogeneous HBM structures per trajectory: \emph{paged KV-cache radix blocks} and \emph{non-sliceable Mamba recurrent state slots}. Under memory pressure, LRU eviction drops long-context \texttt{7/8} stragglers waiting on sandbox returns, forcing costly recomputation. \sys{} injects $\Pi(i)$ into both KV and Mamba evictors (Figure~\ref{fig:l2-scheduling}a, right) to approximate the minimum-sensitivity eviction set $\mathcal{E}^*$:
\begin{equation}
  \mathcal{E}^* = \arg\min_{\mathcal{E} \subseteq \mathcal{A}_{\mathrm{mem}}} \sum_{i \in \mathcal{E}} \Pi(i),
  \quad\text{s.t.}\;\;
  \sum_{i \in \mathcal{E}} m_i^{k} \ge \Delta M_{k},
  \label{eq:co-eviction}
\end{equation}
across $k \in \{\mathrm{KV}, \mathrm{Mamba}\}$ via greedy ascending-$\Pi(i)$ paired eviction: high-completeness groups ($c_g/K \ge 0.5$, i.e., \texttt{4/8--7/8}) are pinned, while low-completeness groups ($c_g/K < 0.5$, i.e., \texttt{$\le$3/8}) are co-evicted in matched KV+Mamba pairs until both deficits $\Delta M_k$ clear.

\paragraph{3. Decode-pressure-aware cross-instance adaptive prefill batching.}
Under Expert Parallelism with Data-Parallel Attention (EP + DP-Attention), replicas (\texttt{DP0--DP3}) share synchronous MoE All-to-All barriers: even with intra-replica chunked prefill~\cite{agrawal2024sarathi}, an unaligned prefill chunk on any replica forces peer decode replicas to stall across heavier MoE expert dispatch/combine and KDA/Mamba state-scan kernels (\texttt{EP/DP Sync Wait}, Figure~\ref{fig:l2-scheduling}b). \sys{} monitors high-priority decode pressure $D_{\mathrm{high}}(t) = \sum_{i \in \mathcal{A}_{\mathrm{dec}}} \mathbb{I}(c_{g(i)}/K > 0.5)$ across replicas and switches $(B_{\mathrm{pre}}(t), \Delta\tau_{\mathrm{pre}}(t))$ between two operating regimes around threshold $\theta_{\mathrm{dec}}$ (Figure~\ref{fig:l2-scheduling}c):
\begin{equation}
  \big(B_{\mathrm{pre}}(t),\, \Delta\tau_{\mathrm{pre}}(t)\big) =
  \begin{cases}
    (B_{\max},\, \tau_{\max}), & D_{\mathrm{high}}(t) \ge \theta_{\mathrm{dec}}, \\
    (B_{\min},\, 0),           & D_{\mathrm{high}}(t) < \theta_{\mathrm{dec}}.
  \end{cases}
\label{eq:prefill-batch}
\end{equation}
In the \emph{Decode-Heavy regime} ($D_{\mathrm{high}}(t) \ge \theta_{\mathrm{dec}}$, e.g., $\ge$3 of 4 replicas decoding \texttt{>4/8} groups), \sys{} sets $(B_{\max}, \tau_{\max}) = (16, 500\text{\,ms})$ (4 aligned prefills/replica) to amortize MoE barrier stalls while closing critical groups; in the \emph{Decode-Light regime} (\texttt{$\le$4/8} groups), it drops to $(B_{\min}, 0) = (4, 0\text{\,ms})$ for immediate admission.

\subsection{Resource-Level (L3): Coverage-Driven Training--Rollout Resource Allocation}
\label{sec:method-l3}

\begin{figure}[t!]
  \centering
  \includegraphics[width=\columnwidth]{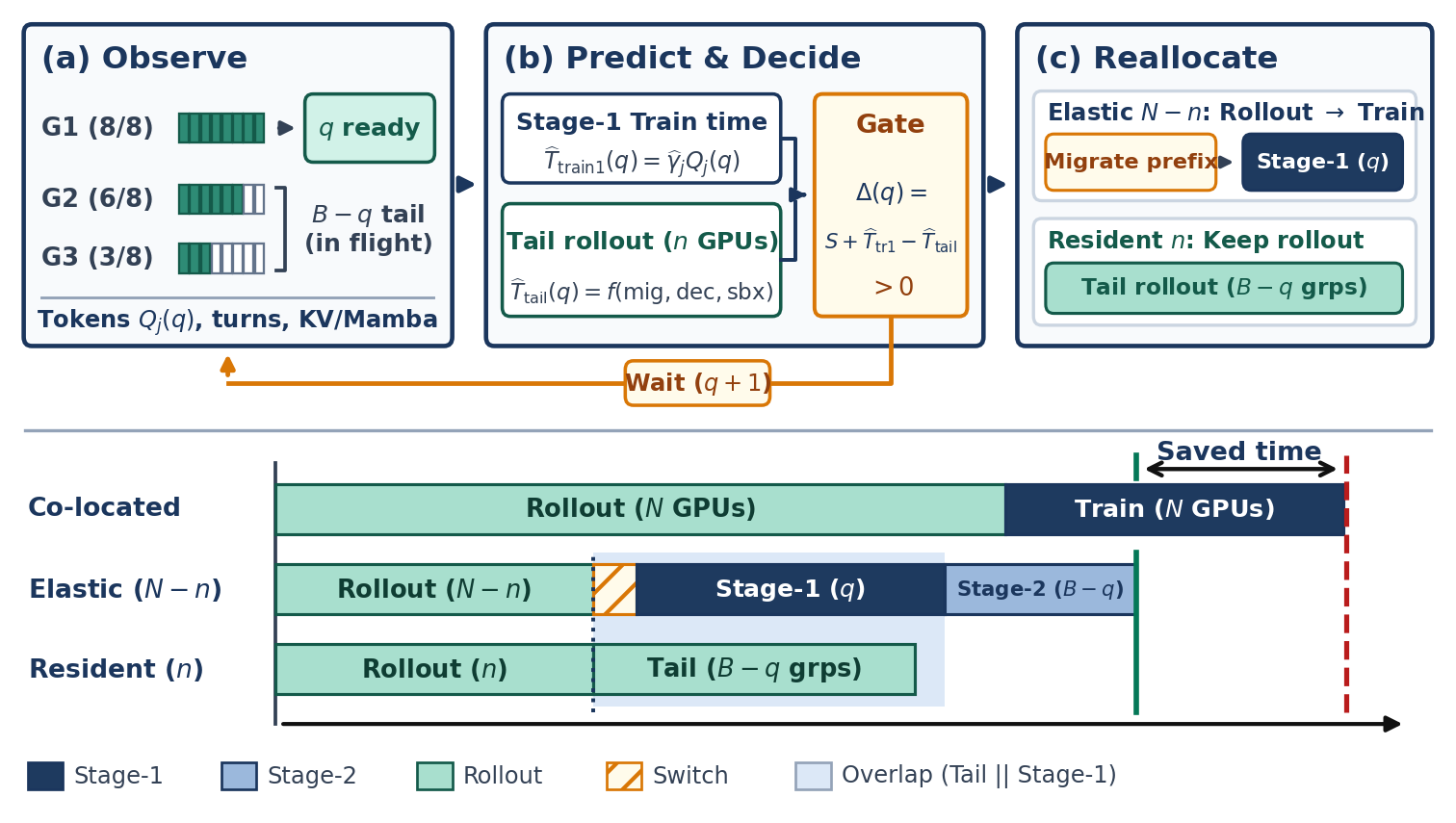}
  \caption{Resource-level (L3) coverage-driven training--rollout resource allocation: closed-loop \texttt{Observe $\to$ Predict \& Decide $\to$ Reallocate} control (top) and phase-overlap timeline (bottom).}
  \label{fig:l3-switching}
\end{figure}

As L2 delivers complete groups into $\mathcal{Q}_j^{\mathrm{ready}}$ and active concurrency tapers toward sandbox-bound stragglers, L3 shifts elastic GPUs from bandwidth-bound low-batch rollout into compute-dense Phase~II Stage-1 training---finishing updates sooner to unlock the next staleness window rather than merely shifting work across steps---via an online \emph{Observe $\to$ Predict \& Decide $\to$ Reallocate} loop (Figure~\ref{fig:l3-switching}).

\paragraph{1. Observe \& Predict: Dual-side duration modeling.}
Whenever a completed group increments $q = |\mathcal{Q}_j^{\mathrm{ready}}|$, L3 reads $s_t$ (Figure~\ref{fig:l3-switching}a) and predicts both post-switch durations (Figure~\ref{fig:l3-switching}b):
\begin{itemize}[leftmargin=*, topsep=2pt, itemsep=1.5pt]
  \item \emph{First-stage training duration $\widehat{T}_{\mathrm{train1}}(q)$}: Since all trajectories in the $q$ ready groups are complete, their exact valid token count $Q_j(q) = \sum_{g=1}^{q}\sum_{i=1}^{K} L_{g,i}$ is known. Using an online least-squares per-token forward--backward slope $\widehat{\gamma}_j$ calibrated from prior steps:
  \begin{equation}
    \widehat{T}_{\mathrm{train1}}(q) = \widehat{\gamma}_j\, Q_j(q).
    \label{eq:pred-train1}
  \end{equation}

  \item \emph{Post-switch tail rollout duration $\widehat{T}_{\mathrm{tail}}(q)$}: Switching $N-n$ elastic GPUs leaves $n$ resident GPUs to finish $B-q$ groups from the eligible in-flight candidate pool $\mathcal{C}_j(q)$ ($|\mathcal{C}_j(q)| \ge B-q$, typically $|\mathcal{C}_j(q)| > B-q$ when cross-step in-flight groups can fulfill Step~$j$; Figure~\ref{fig:l3-switching}c), migrating active prefixes (\texttt{tokens} and \texttt{logprobs}) at prefill recovery cost $V_i$. For each unfinished trajectory $i \in \mathcal{U}_g$ ($g \in \mathcal{C}_j(q)$), L3 reads its elapsed phase time and EMA-tracked historical turn/tool means conditioned on turn index, separating remaining delay into \emph{resident-queued GPU-decode mean $\mu_i$} (scaled by $(N/n)$ congestion) and \emph{GPU-insensitive sandbox wait $\nu_i$} (plus $V_i$) under a shifted exponential completion CDF $\widehat{F}_i(T) = 1 - \exp(-(T - \tau_{i,\mathrm{start}})_+ / (\mu_i + \nu_i))$. Aggregating $\widehat{p}_g(T) = \prod_{i \in \mathcal{U}_g} \widehat{F}_i(T)$ over $\mathcal{C}_j(q)$ yields the expected supply curve $m_j(T; q)$ and tail estimate $\widehat{T}_{\mathrm{tail}}(q)$ at coverage quantile $\eta \in (0,1)$ (default $\eta = 0.88$, ensuring finite crossing even when $|\mathcal{C}_j(q)| = B-q$) with online additive residual margin $\varepsilon_j$:
  \begin{align}
    m_j(T; q) &= \sum_{g \in \mathcal{C}_j(q)} \widehat{p}_g(T)
    = \sum_{g \in \mathcal{C}_j(q)} \prod_{i \in \mathcal{U}_g} \widehat{F}_i(T),
    \label{eq:supply-curve} \\
    \widehat{T}_{\mathrm{tail}}(q) &= \inf\!\big\{T \ge 0 \;\big|\; m_j(T; q) \ge \eta(B - q)\big\} + \varepsilon_j.
    \label{eq:pred-tail}
  \end{align}
\end{itemize}

\paragraph{2. Gate \& Reallocate: Online coverage-margin reallocation.}
Within the feasible switching window $[q_{\min}, q_{\max}]$ (default $q_{\min} = 0.5B$, fallback cap $q_{\max} = 0.75B$), L3 evaluates the online \emph{coverage margin}:
\begin{equation}
  \Delta(q) = S(q) + \widehat{T}_{\mathrm{train1}}(q) - \widehat{T}_{\mathrm{tail}}(q).
  \label{eq:coverage-margin}
\end{equation}
When $\Delta(q) \le 0$ and $q < q_{\max}$ (\emph{heavy tail load}), ready groups cannot mask post-switch tail rollout, so L3 retains all $N$ GPUs in Full Rollout (\texttt{Wait (q+1)}); at $q_j^* = \min\big(\{q \ge q_{\min} \mid \Delta(q) > 0\} \cup \{q_{\max}\}\big)$, L3 drains the $N-n$ elastic GPUs, migrates $\mathcal{F}_{\mathrm{cur}}$ and high-$c_g$ prefixes to the $n$ resident GPUs, and launches \texttt{Stage-1 Train} on $q_j^*$ groups overlapping tail rollout on the remaining $B-q_j^*$ groups, followed by \texttt{Stage-2 Train}. After each step, $\widehat{\gamma}_{j+1}$ updates via exponentially weighted least squares on $(Q_j(q_j^*), T_{\mathrm{train1}})$ and $\varepsilon_{j+1} \leftarrow (1-\alpha)\varepsilon_j + \alpha\big[T_{\mathrm{tail}} - \widehat{T}_{\mathrm{tail}}^{\mathrm{raw}}\big]$ ($\alpha=0.2$) adapts residual tail bias when uncovered wait occurs. Algorithm~\ref{alg:l3} summarizes the complete three-level control loop.

\begin{algorithm}[t!]
\caption{\sys{} Three-Level Co-Orchestration Control Loop (L1 + L2 + L3)}
\label{alg:l3}
\label{alg:l1-l2}
\footnotesize
\begin{algorithmic}[1]
\Require Batch size $B$; bounds $(q_{\min}, q_{\max})$; GPUs $(N, n)$; state $s_t$; parameters $(\widehat{\gamma}_j, \varepsilon_j)$.
\State \textbf{Init Step $j$}: Set all $N$ GPUs to \textbf{Phase~I (Full Rollout)}; $\texttt{switched} \leftarrow \text{False}$.
\LineComment{L1: Water-Level-Gated Dynamic Slot Admission (\S\ref{sec:method-l1})}
\State \textbf{on} \texttt{SlotReleased}($g, r$): update $(c_g, a_g)$; rank $\mathcal{F}_{\mathrm{cur}}$ \& warm sticky continuations (by $c_{g'}\!\downarrow$) $\succ \mathcal{F}_{\mathrm{next}}$; admit onto sticky or least-loaded DP rank satisfying physical KV water-level Eq.~\eqref{eq:admission-gate} ($(1-\rho)C_r^{\mathrm{KV}}$ cap on $c_g=0$ prefetch).
\LineComment{L2: Group-Priority Spatio-Temporal Scheduling (\S\ref{sec:method-l2})}
\State Score active trajectories $\Pi(i)$ via Eq.~\eqref{eq:priority-score}; order replica queues by $\Pi(i)\!\downarrow$; on HBM cap, pin $c_g/K \ge 0.5$ states and greedily co-evict $\mathcal{E}^*$ from KV \& Mamba pools via Eq.~\eqref{eq:co-eviction}; batch prefills via $(B_{\mathrm{pre}}(t), \Delta\tau_{\mathrm{pre}}(t))$ in Eq.~\eqref{eq:prefill-batch}.
\LineComment{L3: Coverage-Driven Training--Rollout Resource Allocation (\S\ref{sec:method-l3})}
\State \textbf{on} \texttt{GroupCompleted}($g$) arriving at $\mathcal{Q}_j^{\mathrm{ready}}$ ($q \leftarrow |\mathcal{Q}_j^{\mathrm{ready}}|$) \textbf{do}
  \State \quad \textbf{if} $\neg\,\texttt{switched}$ \textbf{and} $q_{\min} \le q \le q_{\max}$ \textbf{then}
    \State \qquad Compute $\widehat{T}_{\mathrm{train1}}(q) \leftarrow \widehat{\gamma}_j Q_j(q)$ (Eq.~\eqref{eq:pred-train1}) and $\widehat{T}_{\mathrm{tail}}(q)$ (Eqs.~\eqref{eq:supply-curve}--\eqref{eq:pred-tail});
    \State \qquad \textbf{if} $\Delta(q) = S(q) + \widehat{T}_{\mathrm{train1}}(q) - \widehat{T}_{\mathrm{tail}}(q) > 0$ \textbf{or} $q = q_{\max}$ \textbf{then}
      \State \qquad\quad $\texttt{switched} \leftarrow \text{True}$; $q_j^* \leftarrow q$; migrate $\mathcal{F}_{\mathrm{cur}}$ \& high-$c_g$ prefixes to $n$ resident GPUs and park low-$c_g$ prefetches;
      \State \qquad\quad \textbf{Phase~II}: $N-n$ GPUs run \textbf{Stage-1 Train} on $q_j^*$ groups while $n$ GPUs run \textbf{Tail Rollout} on remaining $B-q_j^*$ groups;
    \State \qquad \textbf{end if}
  \State \quad \textbf{end if}
\State \textbf{when} all $B$ groups ready \textbf{and} Stage-1 Train done: run \textbf{Stage-2 Train} on $B-q_j^*$ groups, update \& broadcast weights (\textbf{Phase~III}), and update $(\widehat{\gamma}_{j+1}, \varepsilon_{j+1})$ via EMA.
\end{algorithmic}
\end{algorithm}

\section{System Implementation}
\label{sec:impl}

We implemented \sys{} in roughly 12K lines of Python atop our distributed RL post-training stack, integrating SGLang~\cite{zheng2024sglang} for inference, Megatron-LM~\cite{shoeybi2019megatron,narayanan2021megatron} for 3D-parallel training, and three open-source infrastructure components: ARouter~\cite{ling2026ling26} for request routing and cross-replica continuation, TransferQueue~\cite{han2025asyncflow} for group-level trajectory aggregation and streaming dispatch, and AState~\cite{inclusionai2025astate} for high-speed cross-pool weight synchronization.

\paragraph{Non-blocking sandbox I/O and L1 admission control plane.}
Each multi-turn trajectory runs in an \texttt{asyncio} coroutine that offloads containerized sandbox executions (shell commands, repository edits, and unit tests) to a CPU I/O thread pool so external stalls never block GPU inference. An asynchronous telemetry collector between \texttt{ARouter} and each SGLang replica tracks per-DP-Attention KV token occupancy (\texttt{used\_tokens/capacity\_tokens}), active running counts (\texttt{Running}), and warm session-to-replica bindings. On turn completion or slot release, \texttt{ARouter} triggers L1 to admit current-step and warm sticky continuations immediately while gating brand-new cross-step prefetches ($c_g=0$) against the physical KV water-level in Eq.~\eqref{eq:admission-gate} ($1-\rho=85\%$) and routing first-turn requests to the least-loaded DP rank; during elastic role switching, L1 migrates $\mathcal{F}_{\mathrm{cur}}$ and high-$c_g$ trajectories ($c_g/K \ge 0.5$) within resident headroom and parks low-completeness prefetches ($c_g/K < 0.5$) at inter-turn boundaries (where sandbox commands have already returned and containers sit idle without triggering timeouts).

\paragraph{Group-aware SGLang scheduling and hybrid state co-protection.}
We extended SGLang's Request Control Block with \texttt{group\_id} and barrier-sensitivity priority $\Pi(i)$ to order local \texttt{waiting} queues. For hybrid MLA + KDA/Mamba models, we unified eviction hooks across SGLang's paged radix tree (\texttt{RadixCache}) and recurrent state allocator (\texttt{Mamba State Pool}), pinning KV pages and Mamba slots for high-completeness groups ($c_g/K \ge 0.5$, \texttt{4/8--7/8}) while greedily co-evicting low-completeness groups ($c_g/K < 0.5$, \texttt{$\le$3/8}) in matched pairs. Within each EP + DP-Attention group, a cross-replica prefill micro-batch buffer ahead of the MoE All-to-All barrier adapts $(B_{\mathrm{pre}}, \Delta\tau_{\mathrm{pre}})$ to live high-priority decode load.

\paragraph{Fidelity-preserving prefix continuation and switching overhead.}
Upon L3 switching, elastic-pool SGLang replicas enter \texttt{Drain} mode: an extended preemption-export hook halts migrating requests and exports generated token IDs with exact behavior-policy log-probabilities (\texttt{rollout\_logprobs}, $\sim$hundreds of KB/trajectory), which \texttt{ARouter} reroutes to resident replicas for fast FP8 prefill continuation ($\sim$0.1--0.3\,s/trajectory on H200). Compared with transferring multi-GB MLA KV caches and resharding non-sliceable KDA/Mamba recurrent states across changing TP/EP topologies over RDMA, token-level continuation avoids cross-pool state-resharding stalls while L1's selective quota bounds migrated prefill load on the $n$ resident GPUs and preserves exact importance-sampling ratios (\S\ref{sec:eval-fidelity}). Once drained, PyTorch CUDA virtual memory APIs release SGLang memory pools and offload inference weights to pinned host DRAM, restoring Megatron-LM shards and gradient buffers in place while keeping NCCL communicators alive. The elastic trainer streams the first $q_j^*$ groups from \texttt{TransferQueue} for Stage-1 1F1B accumulation, processes the trailing $B-q_j^*$ groups in Stage~2, runs a single optimizer step, and broadcasts weights zero-copy via \texttt{AState} over NVLink/RDMA. On 64 H200 GPUs (Figure~\ref{fig:e2e-perf}), mid-step elastic drain and CUDA VMM handoff $S(q)$ take $\sim$1.5\,s (overlapped with resident tail rollout) and the Muon optimizer step takes 1.4\,s (included in the 70\,s \texttt{Train} bar), while the 26.2\,s \texttt{Sync \& Ovh} bar comprises 25.8\,s for end-of-step role restoration and \texttt{AState} weight broadcast plus 0.4\,s for advantage normalization and checkpoint bookkeeping (matching co-located \texttt{veRL}'s 23.6\,s sync cost); the vectorized NumPy L3 predictor itself takes $<$2\,ms per invocation on an asynchronous CPU thread.

\section{Evaluation}
\label{sec:eval}

We evaluate \sys{} in terms of end-to-end training efficiency (\S\ref{sec:eval-e2e}), layer-by-layer ablations of L1--L3 (\S\ref{sec:eval-l1}--\S\ref{sec:eval-l3}), and long-run convergence fidelity (\S\ref{sec:eval-fidelity}).

\subsection{Experimental Setup}
\label{sec:eval-setup}

\paragraph{Cluster, model, and workloads.}
Experiments run on a 64-GPU NVIDIA H200 cluster (144\,GB HBM/GPU, intra-node NVLink, and inter-node RoCEv2 RDMA), evaluating the open-source \texttt{Ling-3.0-Flash}~\cite{inclusionai2026ling3flash} (124B FP8 hybrid sparse MoE combining MLA~\cite{deepseek2024v2} with KDA/Mamba linear-attention caching~\cite{kimi2025linear,gu2024mamba}) on a multi-turn software-engineering agent workload (\texttt{SWE-Agent}), where agents iteratively invoke shell commands, edit code, and run unit tests inside containerized sandboxes (up to 96 turns, 128K-token maximum context). To evaluate all seven baselines and ablations under a uniform benchmark budget while stressing multi-step overlap and KV contention, our controlled end-to-end and L2/L3 ablation runs (\S\ref{sec:eval-e2e}, \S\ref{sec:eval-l2}--\S\ref{sec:eval-l3}) execute a 64K-capped task subset ($B=32$ groups/step, $\sim$8.0K mean sequence length, $\sim$2.05M tokens/step) under an aggressive four-batch window (\texttt{offpolicy=4}), whereas our concurrency-sweep (\S\ref{sec:eval-l1}) and 200-step long-run convergence runs (\S\ref{sec:eval-fidelity}) operate at the full 128K-token context ($B=64$ groups/step, $\sim$50K--58.9K mean sequence length).
Convergence accuracy (\S\ref{sec:eval-fidelity}) is evaluated on SWE-ML~\cite{yang2025swesmith} and SWE-Pro~\cite{deng2026swebenchpro}.

\paragraph{RL configuration and parallelism.}
All runs use GRPO~\cite{shao2024deepseekmath} with $K=8$ trajectories/group (256 trajectories/step at $B=32$ in controlled benchmarks; 512 trajectories/step at $B=64$ in full-context runs) and parameterize asynchronous concurrency by the active in-flight batch window $\mathrm{offpolicy} \in \{1, 2, 3, 4\}$ (allowing up to $\mathrm{offpolicy}-1$ cross-step lookahead steps, so $\mathrm{offpolicy}=1$ is single-step synchronous and $\mathrm{offpolicy}=2$ bounds staleness to $1.0$ step). SGLang runs rollout with $\mathrm{TP}=4$, $\mathrm{DP\text{-}Attention}=4$, $\mathrm{EP}=4$ (DeepEP + DeepGEMM), and Megatron runs 3D-parallel training~\cite{narayanan2021megatron} ($\mathrm{TP}=2$, $\mathrm{PP}=4$, $\mathrm{EP}=8$) with the Muon optimizer~\cite{jordan2024muon,liu2025muon} ($\mathrm{lr}=7.5 \times 10^{-6}$) and AState weight synchronization.
Elastic systems partition the 64-GPU cluster symmetrically into 32 elastic (32E) + 32 resident (32R) GPUs.

\paragraph{Baselines, ablations, and metrics.}
On the same Megatron+SGLang substrate, we compare three architectural baselines alongside full \sys{} and its layer-by-layer ablations:
(1)~\emph{\texttt{veRL}} (64G co-located)~\cite{sheng2025hybridflow}, alternating all GPUs between rollout and training---run in synchronous on-policy mode (\texttt{offpolicy=1}) in \S\ref{sec:eval-e2e} to isolate co-located barrier overhead, and with asynchronous cross-step rollout carry-over (\texttt{offpolicy=2}, $1.0$-step staleness bound) in \S\ref{sec:eval-fidelity} to match our long-run convergence setting;
(2)~\emph{\texttt{AReaL}} (32T+32R static disaggregated async, \texttt{offpolicy=4})~\cite{fu2025areal};
(3)~\emph{\texttt{QwenGyre}}~\cite{wang2026qwengyre}: because \texttt{QwenGyre} is not publicly open-sourced, we faithfully re-implement its elastic reallocation and streaming 1F1B training mechanism on our unified substrate---transitioning elastic cells into streaming training once remaining rollout slots cover unfinished executions ($w(t) \le C_R(t)-C_e$) and ready groups reach preset ratio $\rho_{\min}=62.5\%$, under slot-capped \texttt{offpolicy=4} and per-request FCFS rollout (at 64 GPUs, 32E forms a single $\mathrm{TP}\!\times\!\mathrm{PP}\!\times\!\mathrm{EP}=32$ training cell; we omit its orthogonal branching-harness deduplicator as all systems run standard \texttt{SWE-Agent} trajectories); and
(4--6)~\emph{\texttt{w/o L1}} (L2+L3), \emph{\texttt{w/o L2}} (L1+L3), and \emph{\texttt{w/o L3}} (L1+L2 paired with \texttt{QwenGyre}'s slot-waterlevel and $\rho_{\min}=62.5\%$ trigger to isolate L3's dual-side coverage prediction).
Over post-warmup steady-state windows, we report Mean Step E2E Time, Exposed Rollout Time (Phase~I rollout plus uncovered post-switch tail wait), Normalized Throughput (relative end-to-end speedup), KV Cache Evictions/Step, and converged benchmark accuracy.

\subsection{Overall Performance and Resource Efficiency}
\label{sec:eval-e2e}

\begin{figure}[t!]
  \centering
  \includegraphics[width=\columnwidth]{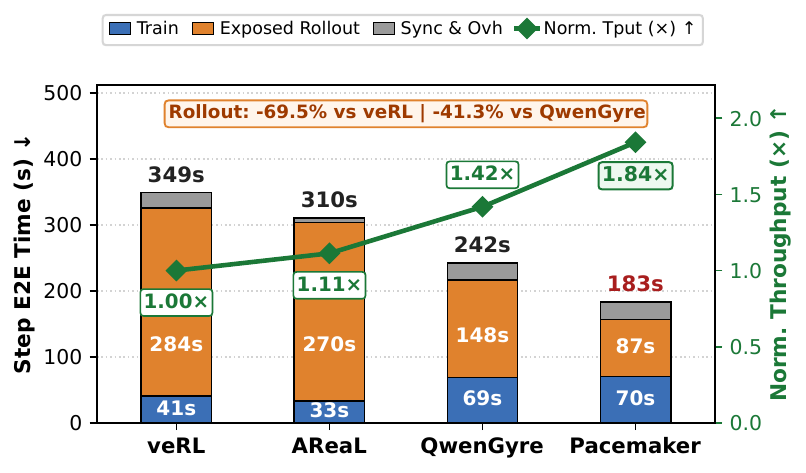}
  \caption{End-to-end step-time breakdown and normalized throughput on \texttt{Ling-3.0-Flash} (124B, 64 H200 GPUs).}
  \label{fig:e2e-perf}
\end{figure}

As shown in Figure~\ref{fig:e2e-perf}, \sys{} achieves \textbf{1.84$\times$}, \textbf{1.66$\times$}, and \textbf{1.30$\times$} higher normalized end-to-end throughput than \texttt{veRL}, \texttt{AReaL}, and \texttt{QwenGyre}, respectively, driven by a \textbf{41.3\%--69.5\%} reduction in exposed rollout wait. Note that the 1.84$\times$ speedup over synchronous \texttt{veRL} (\texttt{offpolicy=1}) reflects the joint benefit of cross-step pipelining and co-orchestration, whereas the 1.66$\times$ and 1.30$\times$ gains over \texttt{AReaL} and \texttt{QwenGyre} under matched \texttt{offpolicy=4} (and 1.31$\times$ over async \texttt{veRL} at \texttt{offpolicy=2} in \S\ref{sec:eval-fidelity}) isolate \sys{}'s orchestration gains.
Each baseline exhibits a distinct architectural bottleneck.
\texttt{veRL} maximizes training throughput by utilizing all 64 GPUs, but full-batch synchronization blocks the entire cluster on each step's slowest long-horizon trajectory, leaving over 80\% of step time in exposed rollout.
\texttt{AReaL} overlaps rollout and training across steps, yet permanently splitting the cluster halves rollout compute and KV capacity; under long-context bursts, severe KV thrashing ($3.1\times$ more evictions than \sys{}) prevents rollout from keeping pace with training.
\texttt{QwenGyre} recovers full-cluster rollout capacity via elastic borrowing, but KV-ungated (slot-only) off-policy prefetching ($2.8\times$ more evictions) and inference-count-driven transitions ($\rho_{\min}=62.5\%$, oblivious to actual training vs.\ tail-rollout durations) still leave heavy-tail groups exposed on the shrunk resident pool.
By coordinating KV-aware admission (L1), group-completeness scheduling (L2), and coverage-driven resource allocation (L3), \sys{} eliminates these rollout stalls and sustains high throughput across diverse multi-turn agent trajectories.

\subsection{Workload-Level (L1) Analysis: Water-Level-Gated Dynamic Request Admission}
\label{sec:eval-l1}

\begin{figure}[t!]
  \centering
  \includegraphics[width=\columnwidth]{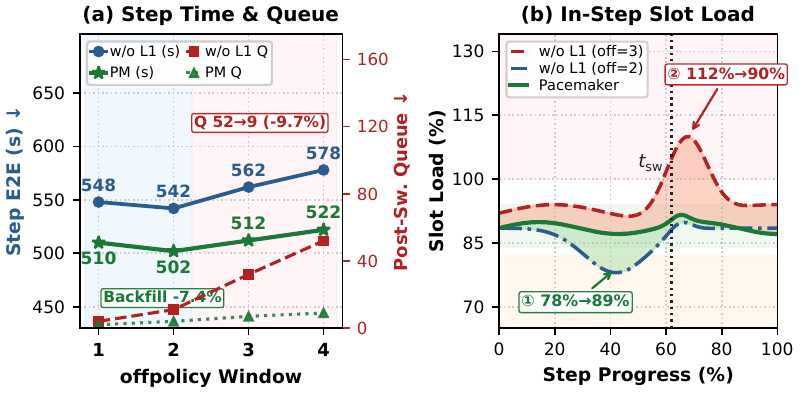}
  \caption{L1 admission ablation across $\mathrm{offpolicy} \in \{1, 2, 3, 4\}$ (\texttt{Ling-3.0-Flash}, 64 H200 GPUs, 128K context): (a)~step time and post-switch resident queue; (b)~in-step pool slot load trajectory.}
  \label{fig:l1-eval}
\end{figure}

Holding L2 and L3 active, Figure~\ref{fig:l1-eval} evaluates L1 across concurrency levels $\mathrm{offpolicy} \in \{1, 2, 3, 4\}$ at the full 128K-token context window (64 H200 GPUs, 32E+32R), isolating both slot-release replenishment and water-level-gated selective migration. Across all four off-policy windows, L1 consistently reduces step time by \textbf{6.9\%--9.7\%} (502--522\,s vs.\ 542--578\,s for \texttt{w/o L1}). Under low concurrency ($\mathrm{offpolicy} \in \{1, 2\}$), group-level batch reservations in \texttt{w/o L1} hold freed slots idle until intra-group stragglers finish; L1's slot-release wakeup immediately backfills freed capacity with prioritized candidates, cutting step time by \textbf{6.9\%--7.4\%} (reaching 502\,s at $\mathrm{offpolicy}=2$). Under high concurrency ($\mathrm{offpolicy} \in \{3, 4\}$), \texttt{w/o L1} admits cross-step prefetches without KV gating and drains displaced elastic-pool traffic onto the halved resident pool at $t_{\mathrm{sw}}$, driving post-switch queues to 32--52 requests; by gating brand-new ($c_g=0$) lookahead admissions at the 85\% physical KV water-level and selectively migrating only high-$c_g$ groups, L1 caps resident queues at 7--9 requests and cuts step time by \textbf{8.9\%--9.7\%}.

Figure~\ref{fig:l1-eval}b traces the in-step slot demand-to-capacity ratio (active plus queued requests normalized by pool slot capacity) to illustrate how L1 avoids both static-window failure modes. Without L1, static low concurrency ($\mathrm{offpolicy}=2$) suffers a mid-step vacancy trough dipping to 78\%, whereas static higher concurrency ($\mathrm{offpolicy}=3$) overshoots capacity to a $\sim$110\%--112\% post-switch queueing spike after $t_{\mathrm{sw}}$. By dynamically backfilling freed slots and gating cross-step prefetches, L1 eliminates both extremes simultaneously---lifting mid-step occupancy to 89\% (vs.\ 78\% under \texttt{w/o L1}, $\mathrm{offpolicy}=2$) while holding post-switch slot load near 90\% (vs.\ 110\%--112\% under \texttt{w/o L1}, $\mathrm{offpolicy}=3$). On the 64K-context benchmark at $\mathrm{offpolicy}=4$, L1 exhibits the same memory-stabilizing behavior, cutting steady-state KV cache evictions by \textbf{54.6\%} (and reducing exposed rollout by 6.0\%).

\subsection{Runtime-Level (L2) Analysis: Group-Priority Spatio-Temporal Adaptive Scheduling}
\label{sec:eval-l2}

\begin{figure}[t!]
  \centering
  \includegraphics[width=\columnwidth]{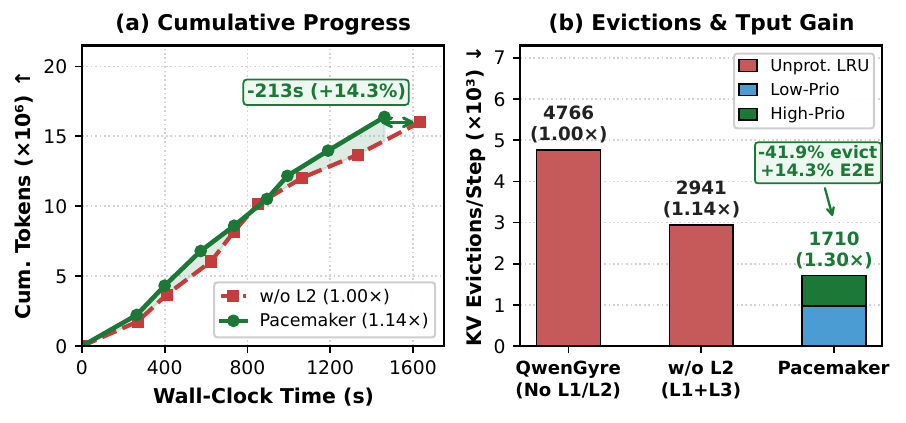}
  \caption{L2 scheduling ablation (\texttt{w/o L2} vs.\ \sys{}): (a)~cumulative training tokens delivered versus wall-clock time; (b)~KV eviction control with low- vs.\ high-priority separation and E2E throughput uplift.}
  \label{fig:l2-eval}
\end{figure}

With L1 and L3 active, Figure~\ref{fig:l2-eval} compares \sys{} against \texttt{w/o L2} (standard FCFS queueing, LRU eviction, and immediate prefill). Under FCFS and immediate prefill, bursts of newly admitted trajectories interleave long prefills ahead of decode steps belonging to nearly complete groups, creating head-of-line blocking that delays the step barrier. By continuously reordering ready queues according to group-completeness priority ($\sim$19.6K queue reorders/step) and deferring low-priority prefills ($\sim$224 deferrals/step), L2 prevents closing stragglers from being starved by fresh arrivals. In parallel, Figure~\ref{fig:l2-eval}b contrasts unprotected LRU eviction in \texttt{QwenGyre} and \texttt{w/o L2} against L2's barrier-aware hybrid KV/Mamba co-eviction: even after L1 trims indiscriminate evictions from \texttt{QwenGyre} to \texttt{w/o L2}, progress-oblivious LRU still evicts trajectories paused in sandbox execution and triggers costly $O(T^2)$ prefix recomputation. Aligning eviction order with barrier sensitivity and co-evicting KV and Mamba states together cuts evictions by a further \textbf{41.9\%} over \texttt{w/o L2} and shifts \textbf{56.8\%} of remaining evictions into the low-priority prefetch band rather than high-priority closing trajectories.

Figure~\ref{fig:l2-eval}a shows how eliminating both head-of-line blocking and repeated prefix recomputation translates into sustained end-to-end progress across the steady-state window: \sys{} stays ahead of \texttt{w/o L2} across nearly the entire window, delivering 16M training tokens 213\,s earlier while lowering worst-step latency from 300.7\,s to 272.5\,s. Overall, L2 cuts exposed rollout wait by \textbf{19.2\%}, shortens mean step time by \textbf{10.5\%}, and lifts end-to-end throughput by \textbf{+14.3\%} (Figure~\ref{fig:l2-eval}b), confirming that engine-level group-priority scheduling is essential even when L1 and L3 are active.

\subsection{Resource-Level (L3) Analysis: Coverage-Driven Training--Rollout Resource Allocation}
\label{sec:eval-l3}

\begin{figure}[t!]
  \centering
  \includegraphics[width=\columnwidth]{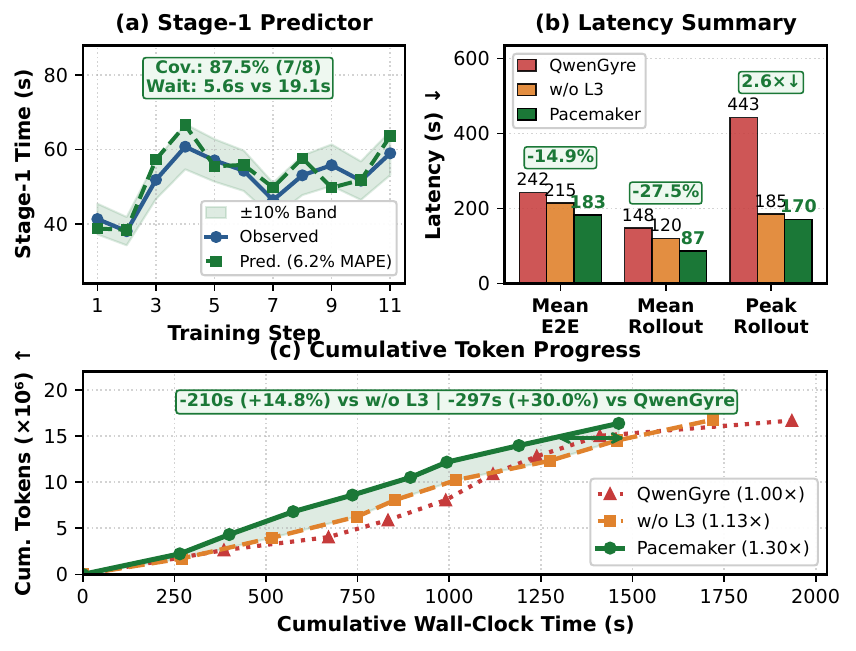}
  \caption{L3 coverage-driven resource allocation comparing \texttt{QwenGyre} ($\rho_{\min}=62.5\%$), \texttt{w/o L3} (L1+L2, $\rho_{\min}=62.5\%$), and \sys{}: (a)~online Stage-1 predictor accuracy (6.2\% MAPE) and post-switch tail coverage across Steps~1--11; (b)~mean step time, exposed rollout, and peak exposed rollout summary; (c)~cumulative training tokens delivered versus wall-clock time.}
  \label{fig:l3-eval}
\end{figure}

With L1 and L2 active, Figure~\ref{fig:l3-eval} evaluates L3 against \texttt{w/o L3} (L1+L2 with \texttt{QwenGyre}'s slot-waterlevel and preset $\rho_{\min}=62.5\%$ trigger) and \texttt{QwenGyre} ($\rho_{\min}=62.5\%$ without L1+L2). Triggering elastic transitions solely from active slot counts and a preset $\rho_{\min}=62.5\%$ supply ratio shifts unfinished tail groups onto the shrunk resident pool regardless of their actual remaining decode and tool latency, inducing resident-pool congestion and uncovered post-switch trainer stalls on heavy-tail steps even with L1 and L2 enabled. To align reallocation with true supply--demand durations without offline profiling, L3 pairs live per-turn tool latency and remaining decode estimates ($\widehat{T}_{\mathrm{tail}}(q)$) with an online linear Stage-1 training predictor $\widehat{T}_{\mathrm{train1}}(q)$---deferring reallocation to the $q_{\max}=75\%$ cap on decode-heavy steps while reallocating early at $62.5\%$ whenever $S(q)+\widehat{T}_{\mathrm{train1}}(q)$ covers the remaining tail. As shown in Figure~\ref{fig:l3-eval}a, $\widehat{T}_{\mathrm{train1}}(q)$ tracks observed Megatron Stage-1 durations across Steps~1--11 with \textbf{6.2\% MAPE}; on the tail side (where successful coverage right-censors tail completion inside Stage-1 training), L3 achieves full post-switch tail coverage ($\le 2.1\text{\,s}$ wait) in \textbf{87.5\% (7/8)} of steady-state steps and cuts mean uncovered post-switch trainer wait to \textbf{5.6\,s}, compared with \textbf{19.1\,s} on \texttt{w/o L3} (62.5\% coverage) and \textbf{99.4\,s} on \texttt{QwenGyre} (25.0\% coverage).

Driven by these real-time coverage estimates, L3 consistently outperforms inference-count-driven reallocation in both latency and throughput (Figure~\ref{fig:l3-eval}b--c). Across the steady-state window (Figure~\ref{fig:l3-eval}b), L3 reduces mean exposed rollout by \textbf{27.5\%} and mean step time by \textbf{14.9\%} relative to \texttt{w/o L3}, while cutting exposed rollout by \textbf{41.3\%} and peak exposed rollout by \textbf{2.6$\times$} relative to \texttt{QwenGyre} (443\,s $\to$ 170\,s). In the cumulative progress trace (Figure~\ref{fig:l3-eval}c), \sys{} widens its lead monotonically across steps with no cross-step catch-up by baselines, delivering 16M training tokens 210\,s faster than \texttt{w/o L3} (\textbf{+14.8\%} throughput) and 297\,s faster than \texttt{QwenGyre} (\textbf{+30.0\%})---confirming that L3 eliminates unrecoverable barrier waits and resident KV evictions (\textbf{$-$20.9\%} beyond L1+L2) rather than borrowing work from future steps.

\begin{figure}[t!]
  \centering
  \includegraphics[width=\columnwidth]{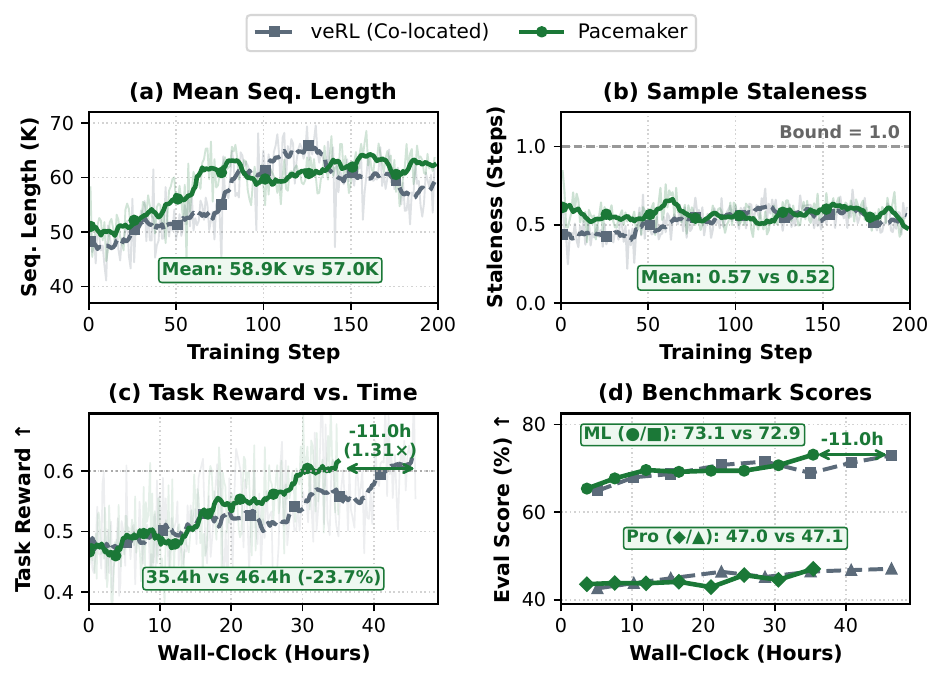}
  \caption{Long-run training fidelity and convergence of \sys{} vs.\ \texttt{veRL} (both at \texttt{offpolicy=2}, $1.0$-step staleness bound) over a 200-step 128K-context \texttt{SWE-Agent} run: (a)~mean sequence length vs.\ step; (b)~sample staleness vs.\ step; (c)~task reward and (d)~SWE-ML / SWE-Pro benchmark scores vs.\ wall-clock hours.}
  \label{fig:fidelity}
\end{figure}

\begin{table}[t!]
  \centering
  \footnotesize
  \caption{Final converged benchmark scores and total wall-clock time (Step~200) on SWE-ML and SWE-Pro for \texttt{Ling-3.0-Flash} (124B).}
  \label{tab:convergence}
  \setlength{\tabcolsep}{4.5pt}
  \begin{tabular}{@{}lccc@{}}
    \toprule
    \textbf{System} & \textbf{SWE-ML} & \textbf{SWE-Pro} & \textbf{200-Step Time} \\
    \midrule
    \texttt{veRL} (Co-located) & 72.92 & \textbf{47.05} & 46.4\,h \\
    \textbf{\sys{}}            & \textbf{73.08} & 47.03 & \textbf{35.4\,h ($-$23.7\%)} \\
    \bottomrule
  \end{tabular}
\end{table}

\subsection{Training Fidelity and End-to-End Convergence}
\label{sec:eval-fidelity}

To verify that \sys{} preserves RL optimization semantics over long-horizon training, Figure~\ref{fig:fidelity} compares \sys{} against co-located \texttt{veRL} across a 200-step 128K-context \texttt{SWE-Agent} run under identical GRPO hyperparameters ($B=64$, $K=8$, $\mathrm{lr}=7.5 \times 10^{-6}$, clip $\epsilon=0.2$) and our standard $1.0$-step staleness bound (\texttt{offpolicy=2}, enabled identically on both \sys{} and \texttt{veRL}'s asynchronous cross-step carry-over). Across all 200 steps, Figure~\ref{fig:fidelity}a shows that both systems follow closely matched sequence-length growth trajectories, expanding naturally from roughly 48--50K tokens to 60--64K tokens as multi-turn reasoning deepens (mean 58.9K for \sys{} vs.\ 57.0K for \texttt{veRL}, $+$3.3\%), confirming that elastic scheduling induces neither length collapse nor runaway verbosity. Likewise, Figure~\ref{fig:fidelity}b shows that \sys{}'s mean sample staleness is only marginally higher than \texttt{veRL} (0.57 vs.\ 0.52 steps, $+$0.05) because overlapping rollout execution with Stage-1 training allows slightly more in-flight tail trajectories to span a single weight-update boundary; nevertheless, staleness remains tightly aligned with \texttt{veRL} and strictly below the $1.0$-step bound throughout training, keeping importance-weighted gradient updates well within safe RL operating bounds.

Figure~\ref{fig:fidelity}c--d and Table~\ref{tab:convergence} confirm that this bounded fidelity translates directly into lossless end-to-end convergence in substantially less wall-clock time. Completing 200 training steps takes 35.4\,hours on \sys{} (mean 637\,s/step, reflecting the 48K$\to$64K sequence-length growth over 200 steps and periodic checkpointing relative to the early-window 502\,s at $\mathrm{offpolicy}=2$ in \S\ref{sec:eval-l1}) versus 46.4\,hours on \texttt{veRL} (\textbf{$-$23.7\%} wall-clock time, \textbf{1.31$\times$} faster, saving 11.0\,hours as avoided tail under-utilization and cross-step re-prefill costs compound across all 200 updates), keeping \sys{}'s task reward consistently ahead on the wall-clock axis from hour~15 onward while converging to matching reward levels (0.604 vs.\ 0.615 late-window mean, Figure~\ref{fig:fidelity}c). Across periodic benchmark evaluations (Figure~\ref{fig:fidelity}d and Table~\ref{tab:convergence}), \sys{} attains virtually identical final accuracy 11.0\,hours earlier---73.08 vs.\ 72.92 on SWE-ML ($+$0.16) and 47.03 vs.\ 47.05 on SWE-Pro ($-$0.02)---demonstrating lossless convergence well within normal evaluation variance.

\section{Related Work}
\label{sec:related}

\paragraph{LLM serving and hybrid state management.}
Continuous batching~\cite{yu2022orca}, PagedAttention~\cite{kwon2023vllm}, and RadixAttention~\cite{zheng2024sglang} maximize request goodput, while Sarathi-Serve~\cite{agrawal2024sarathi} and DistServe~\cite{zhong2024distserve} reduce prefill--decode interference via chunked prefills and P--D disaggregation. Marconi~\cite{pan2025marconi} and Jenga~\cite{zhang2025jenga} further manage hybrid Transformer--SSM memory via FLOP-aware caching and heterogeneous allocation. Unlike online serving, \sys{} optimizes all-$K$ group completion under MoE EP barriers by coupling hybrid KV+Mamba co-eviction with barrier sensitivity and cross-replica prefill batching.

\paragraph{Barrier-aware and RL post-training systems.}
Prioritizing stragglers behind all-$K$ barriers parallels cluster coflow scheduling (Varys~\cite{chowdhury2014varys}, Aalo~\cite{chowdhury2015aalo}), extended here to stateful multi-turn GPU+sandbox execution. Early RLHF systems~\cite{yao2023deepspeedchat,hu2024openrlhf,sheng2025hybridflow,mei2025realhf} target single-turn generation, while asynchronous/streaming frameworks~\cite{zhong2025streamrl,han2025asyncflow,fu2025areal,zhang2026venusrl} and elastic allocators~\cite{gao2026rollpacker,wang2026qwengyre,wang2026dynarl,chen2026libra,tian2026reflow} decouple or shift GPUs via inference-only watermarks (\S\ref{sec:bg-related}); \sys{} closes the loop across admission, hybrid state scheduling, and dual-side coverage-driven reallocation.

\section{Conclusion}
\label{sec:conclusion}

We presented \sys{}, a progress-driven training--rollout co-orchestration framework for long-horizon agentic RL that paces complete-group supply (\texttt{Groups/s}) to match training consumption via water-level-gated admission with selective migration (L1), barrier-sensitive hybrid KV/Mamba scheduling with adaptive prefill batching (L2), and coverage-driven resource allocation (L3). On the open-source 124B \texttt{Ling-3.0-Flash} hybrid MoE model across 64 H200 GPUs, \sys{} cuts exposed rollout wait by 41.3\%--69.5\% ($2.8\times$--$3.1\times$ fewer evictions) to achieve 1.30$\times$--1.84$\times$ end-to-end speedups over \texttt{QwenGyre}, \texttt{AReaL}, and \texttt{veRL} (1.15$\times$ over \texttt{w/o L3}), and shortens 200-step SWE training by 23.7\% (1.31$\times$) with lossless convergence on SWE-ML and SWE-Pro.

\bibliographystyle{ACM-Reference-Format}
\bibliography{references}

\end{document}